\documentclass[sn-mathphys-num]{sn-jnl}

\usepackage{graphicx}\usepackage{multirow}\usepackage{amsmath,amssymb,amsfonts}\usepackage{amsthm}\usepackage{mathrsfs}\usepackage[title]{appendix}\usepackage[x11names]{xcolor}\usepackage{textcomp}\usepackage{array}
\usepackage{manyfoot}\usepackage{booktabs}\usepackage{longtable}
\usepackage{algorithm}\usepackage{algorithmicx}\usepackage{algpseudocode}\usepackage{listings}\usepackage[utf8]{inputenc}
\usepackage{dirtytalk}

\theoremstyle{thmstyleone}

\theoremstyle{thmstyletwo}

\theoremstyle{thmstylethree}

\def\showComments{1}		  \ifnum\showComments=1
	\newcommand{\todo   }[1]{{\leavevmode\color{red}    {[\textbf{TO DO:}     #1]}}}
	\newcommand{\kathrin}[1]{{\leavevmode\color{violet} {[\textbf{Kathrin:}   #1]}}}
	\newcommand{\nick   }[1]{{\leavevmode\color{orange} {[\textbf{Nick:}      #1]}}}
	\newcommand{\fiona  }[1]{{\leavevmode\color{Purple4}{[\textbf{Fiona:}     #1]}}}
	\newcommand{\soumya }[1]{{\leavevmode\color{teal}   {[\textbf{Soumya:}    #1]}}}
	\newcommand{\Boris  }[1]{{\leavevmode\color{purple} {[\textbf{Boris:}     #1]}}}
	\newcommand{\seb    }[1]{{\leavevmode\color{magenta}{[\textbf{Sebastian:} #1]}}}
	\newcommand{\simon  }[1]{{\leavevmode\color{Sienna4}{[\textbf{Simon:}     #1]}}}
\fi
\ifnum\showComments=0
	\newcommand{\todo   }[1]{}
	\newcommand{\kathrin}[1]{}
	\newcommand{\nick   }[1]{}
	\newcommand{\fiona  }[1]{}
	\newcommand{\soumya }[1]{}
	\newcommand{\Boris  }[1]{}
	\newcommand{\seb    }[1]{}
	\newcommand{\simon  }[1]{}
\fi

\newcommand{\heading}[1]{{\vspace{1ex}\noindent\sc{#1}}}

\usepackage{cleveref}

\usepackage[style=numeric-comp,
  sorting=none,
  bibencoding=utf8,
  backend=biber,
  language=auto,    autolang=other
  ]{biblatex} \usepackage{CJKutf8}
\usepackage[utf8]{inputenc}
\usepackage[T2A,T1]{fontenc}
\usepackage[main=english,russian]{babel} \usepackage{textgreek}
\usepackage{csquotes} \usepackage{lmodern} \usepackage{siunitx}
\DeclareSourcemap{ \maps[datatype=bibtex]{
     \map{
        \step[fieldsource=doi,final]
        \step[fieldset=url,null]
        \step[fieldset=urldate,null]
        }
      }
}
\DeclareSourcemap{ \maps[datatype=bibtex]{
     \map{
        \step[fieldsource=doi,final]
        \step[fieldset=howpublished,null]
        }
      }
}
\usepackage[acronym]{glossaries}
\usepackage{glossaries-extra}
\makeglossaries

\usepackage{pdflscape}
\usepackage{makecell}

\begin{document}

\title[A Critical Analysis of Deployed Use Cases for QKD and Comparison with PQC]{A Critical Analysis of Deployed Use Cases for Quantum Key Distribution and Comparison with Post-Quantum Cryptography}

\author[1,3]{\fnm{Nick} \sur{Aquina}}\email{n.aquina@tue.nl}
\equalcont{These authors contributed equally to this work.}

\author[1,3]{\fnm{Bruno} \sur{Cimoli}}\email{b.cimoli@tue.nl}
\equalcont{These authors contributed equally to this work.}

\author[2,3]{\fnm{Soumya} \sur{Das}}\email{s.das2@tue.nl}
\equalcont{These authors contributed equally to this work.} 

\author[2,3]{\fnm{Kathrin} \sur{Hövelmanns}}\email{kathrin@hoevelmanns.net}
\equalcont{These authors contributed equally to this work.}

\author[2,3]{\fnm{Fiona Johanna} \sur{Weber}}\email{crypto@fionajw.de}
\equalcont{These authors contributed equally to this work.}

\author[1,3]{\fnm{Chigo} \sur{Okonkwo}}

\author[1,3]{\fnm{Simon} \sur{Rommel}}

 \author[2,3]{\fnm{Boris} \sur{Škorić}}

\author[1,3]{\fnm{Idelfonso} \sur{Tafur Monroy}}

\author[4]{\fnm{Sebastian} \sur{Verschoor}}

\affil[1]{\orgdiv{Department of Electrical Engineering}, \orgname{Eindhoven University of Technology}, \orgaddress{\street{Flux Building}, \city{Eindhoven}, \postcode{5612 AZ}, \country{The Netherlands}}}

\affil[2]{\orgdiv{Department of Mathematics and Computer Science}, \orgname{Eindhoven University of Technology}, \orgaddress{\street{MetaForum Building}, \city{Eindhoven}, \postcode{5612 AZ}, \country{The Netherlands}}}

\affil[3]{\orgdiv{Eindhoven Quantum Hub}, \orgname{Eindhoven University of Technology}, \orgaddress{\street{Qubit Building}, \city{Eindhoven}, \postcode{5612 AZ}, \country{The Netherlands}}}

\affil[4]{\orgdiv{Informatics Institute}, \orgname{University of Amsterdam}, \orgaddress{\street{Science Park 904}, \postcode{1098 XH} \city{Amsterdam}, \country{The Netherlands}}}

\abstract{Quantum Key Distribution (QKD) is currently being discussed as a technology to safeguard communication in a future where quantum computers compromise traditional public-key cryptosystems. In this paper, we conduct a comprehensive security evaluation of QKD-based solutions, focusing on real-world use cases sourced from academic literature and industry reports. We analyze these use cases, assess their security and identify the possible advantages of deploying QKD-based solutions. We further compare QKD-based solutions with Post-Quantum Cryptography (PQC), the alternative approach to achieving security when quantum computers compromise traditional public-key cryptosystems, evaluating their respective suitability for each scenario. Based on this comparative analysis, we critically discuss and comment on which use cases QKD is suited for, considering factors such as implementation complexity, scalability, and long-term security. Our findings contribute to a better understanding of the role QKD could play in future cryptographic infrastructures and offer guidance to decision-makers considering the deployment of QKD.}

{}


\maketitle

\newacronym{AES}{AES}{Advanced Encryption Standard}
\newacronym{AKE}{AKE}{Authenticated Key Exchange}
\newacronym{ANSSI}{ANSSI}{Agence nationale de la sécurité des systèmes d'information (French)}
\newacronym{AONT}{AONT}{All-or-nothing transform}
\newacronym{ASIC}{ASIC}{Application-Specific Integrated Circuit}
\newacronym{BSI}{BSI}{Bundesamt für Sicherheit in der Informationstechnik (German)}
\newacronym{Ciemat}{Ciemat}{Research Centre for Energy, Environment and Technology}
\newacronym{COW}{COW}{Coherent One Way}
\newacronym{CSIC}{CSIC}{Spanish National Research Council}
\newacronym{CRQC}{CRQC}{Cryptographically Relevant Quantum Computer}
\newacronym{CV-QKD}{CV-QKD}{Continuos Variable QKD}
\newacronym{CWDM}{CWDM}{Coarse Wavelength Division Multiplexing}
\newacronym{DHKX}{DHKX}{Diffie-Hellman Key Exchange}
\newacronym{DM}{DM}{Discrete Modulation}
\newacronym{DPS-QKD}{DPS-QKD}{Differential Phase Shifted QKD}
\newacronym{DRC}{DRC}{Disaster Recovery Center}
\newacronym{DV-QKD}{DV-QKD}{Discrete Variable QKD}
\newacronym{FFT}{FFT}{Fast-Fourier Transform}
\newacronym{FN-DSA}{FN-DSA}{\gls{FFT} over \gls{NTRU}-Lattice-Based Digital Signature Algorithm}
\newacronym{FPGA}{FPGA}{Field-Programmable Gate Array}
\newacronym{GM}{GM}{Guassian Modulation}
\newacronym{GMAC}{GMAC}{Galois Message Authentication Code}
\newacronym{M-LWE}{M-LWE}{Module Learning With Errors}
\newacronym{HSM}{HSM}{Hardware Security Module}
\newacronym{HMAC}{HMAC}{Hash-based Message Authentication Code }
\newacronym{HNDL}{HNDL}{Harvest-Now-Decrypt-Later}
\newacronym{HQC}{HQC}{Hamming Quasi-Cyclic}
\newacronym{IPsec}{IPsec}{Internet Protocol Security}
\newacronym{IRTF}{IRTF}{Internet Research Task Force}
\newacronym{ITS}{ITS}{Information-theoretic security}
\newacronym{LMS}{LMS}{Leighton–Micali Signatures}
\newacronym{LWE}{LWE}{Learning With Errors}
\newacronym{KEM}{KEM}{Key Encapsulation Mechanism}
\newacronym{KM}{KM}{Key Management}
\newacronym{KMS}{KMS}{Key Management System}
\newacronym{MAC}{MAC}{Message Authentication Code}
\newacronym{MACsec}{MACsec}{Media Access Control security}
\newacronym{MAN}{MAN}{Metropolitan Area Network}
\newacronym{MDI-QKD}{MDI-QKD}{Measurement-Device-Independent QKD}
\newacronym{NCSC}{NCSC}{Nationaal Cyber Security Centrum (Dutch)}
\newacronym{NEC}{NEC}{Nippon Electric Company}
\newacronym{NICT}{NICT}{National Institute of Information and Communications Technology}
\newacronym{NIST}{NIST}{National Institute of Standards and Technology (US)}
\newacronym{NTRU}{NTRU}{N-th degree Truncated polynomial Ring Units}
\newacronym{ML-DSA}{ML-DSA}{Module-Lattice-Based Digital Signature Algorithm}
\newacronym{ML-KEM}{ML-KEM}{Module-Lattice-Based Key-Encapsulation Mechanism}
\newacronym[shortplural={OTPs},longplural={One-Time Pads}]{OTP}{OTP}{One-Time Pad}
\newacronym{P2P}{P2P}{Point-to-point}
\newacronym{PKI}{PKI}{Public Key Infrastructure}
\newacronym{PCS}{PCS}{Post-Compromise Security}
\newacronym{PQ}{PQ}{Post-Quantum}
\newacronym{PQC}{PQC}{Post-Quantum Cryptography}
\newacronym{QAM}{QAM}{Quadrature Amplitude Modulation}
\newacronym{Q-ITS}{Q-ITS}{Quantum Information-theoretic security}
\newacronym{QKD}{QKD}{Quantum Key Distribution}
\newacronym{QPSK}{QPSK}{Quadrature Phase Shift Keying}
\newacronym[shortplural={QRNGs},longplural={Quantum Random Number Generators}]{QRNG}{QRNG}{Quantum Random Number Generator}
\newacronym{RSA}{RSA}{Rivest–Shamir–Adleman}
\newacronym{Rx}{Rx}{receiver}
\newacronym{SCADA}{SCADA}{Supervisory Control and Data Acquisition}
\newacronym{SHA}{SHA}{Secure Hash Algorithm}
\newacronym{SIG}{SIG}{Services Industriels de Genève}
\newacronym{SKR}{SKR}{Secret Key Rate}
\newacronym{SLH-DSA}{SLH-DSA}{Stateless Hash-Based Digital Signature Algorithm}
\newacronym{SNR}{SNR}{Signal to Noise Ratio}
\newacronym{SSH}{SSH}{Secure Shell}
\newacronym{SSS}{SSS}{Shamir's Secret Sharing}
\newacronym{TB}{TB}{terabyte}
\newacronym{TLS}{TLS}{Transport Layer Security}
\newacronym{Tx}{Tx}{transmitter}
\newacronym{ToMMo}{ToMMo}{Tohoku Medical Megabank Organization}
\newacronym{UPM}{UPM}{Universidad Politécnica de Madrid}
\newacronym{VPN}{VPN}{Virtual Private Network}
\newacronym{WAN}{WAN}{Wide Area Network}
\newacronym{WDM}{WDM}{Wavelength Division Multiplexing}
\newacronym{XMSS}{XMSS}{eXtended Merkle Signature Scheme}
\newacronym{XOR}{XOR}{Exclusive OR}
 
\section{Introduction}\label{sec1}
	Heavy corporate and public investments are accelerating the development of large-scale quantum computers~\cite{Ichikawa2024,2401.16317}.
This development threatens the confidentiality of current data transmissions because sufficiently powerful quantum computers could break certain cryptographic algorithms~\cite{shorsalgo} that currently are widely used.
Importantly, this is not only a threat to future data, but also to data encrypted today:
in a so-called \glsxtrfull{HNDL} attack, attackers record data on transit and decrypt the data once a quantum computer becomes available, thereby breaching confidentiality retroactively~\cite{prepare-aivd}. While such attacks compromise confidentiality, they do not enable retroactive breaches of authentication (e.g., impersonation), as authentication relies on real-time verification during the interaction.
To protect confidential communication against quantum attacks, it is necessary to overhaul current cryptographic solutions. Currently, the main overhaul strategies in discussion are  a), replacing the broken algorithms with still-classical algorithms that are secure against quantum attacks (so-called \glsxtrfull{PQC}), b), utilize \glsxtrfull{QKD}~\cite{bb84}, and additionally, using mixes of both~\cite{garms2024experimental,zeng2024practical}.

\gls{PQC} essentially allows to maintain already established security infrastructures (with some caveats concerning compatibility), and the first \gls{PQC} solutions have recently been standardized~\cite{NIST_FIPS}.
At the same time, \gls{PQC} is not unconditionally secure and future algorithmic breakthroughs could occur, one thus cannot unconditionally rule out \gls{HNDL} attacks~\cite{EuropeanStatement}, even ones run on classical computers.
While \gls{QKD}-based solutions theoretically overcome this limitation of \gls{PQC}, they will introduce significant changes to security infrastructures and are not yet fully standardized~\cite{cen-cenelec_fgqt_standardization_2023,TSDSI}, though several solutions have been demonstrated at different locations around the world.
\gls{QKD} currently exhibits practical limitations that public-key cryptography does not have -- first, \gls{QKD} is limited in distance and until quantum repeaters~\cite{RevModPhys.95.045006} are developed, \gls{QKD} cannot provide end-to-end security over long distances, and, second, wireless \gls{QKD} technologies currently only exist with free space optics, where photons are sent over the air~\cite{Kri2023}, making it incompatible with the currently predominant radio networks.
These limitations restrict the use cases in which \gls{QKD} can be applied and prevent \gls{QKD} from serving as a general-purpose replacement for public-key cryptography. Additionally, information cannot be protected with \gls{QKD} alone -- to ensure confidentiality and authenticity, it has to be used in conjunction with quantum-secure authentication and quantum-secure symmetric encryption. To assess which level of security can be guaranteed for a concrete use case, it is necessary to account for \textit{all} involved cryptographic algorithms and \textit{all} underlying assumptions. Assessments that account for this fuller picture help clarify the comparison between \gls{QKD}-based solutions and solutions that do not involve \gls{QKD}. This also paves the way towards a comparison between \gls{QKD}-based solutions and solutions stemming from \gls{PQC}, on a case-by-case basis.

Most governmental bodies that have commented on the choice between \gls{QKD} and post-quantum asymmetric cryptography currently prefer the latter:
The US National Security Agency (NSA) considers post-quantum cryptography to be `more cost-effective and easily maintained' and `does not support the use of \gls{QKD}  to protect communications in national security systems'~\cite{USStatement}. Their criticism about \gls{QKD} are addressed in detail in~\cite{2307.15116}.
In a joint position document, the French Cybersecurity Agency (ANSSI), the German Federal Information Security Office (BSI), the Netherlands National Communications Security Agency (NLNCSA) and the Swedish National Communications Security Authority declare `the clear priority should [...] be the migration to post-quantum cryptography' 
and state that \gls{QKD} can `due to current and inherent limitations […] currently only be used in practice in some niche use cases'~\cite{EuropeanStatement}.
However, they also state that `research on this topic should be continued in order to investigate if
there are ways to overcome some of the limitations of the current technology'.
The British National Cyber Security Centre (NCSC) published a statement 
stating that `the NCSC does not endorse the use of \gls{QKD} for any government or military applications, and cautions against sole reliance on \gls{QKD} for business-critical networks' and instead advises `that the best mitigation against the threat of quantum computers is quantum-safe cryptography'. NCSC also recommends any other organizations considering the use of \gls{QKD} as a key agreement mechanism ensure that robust quantum-safe cryptographic mechanisms for authentication are implemented alongside them.'~\cite{UKStatement}.

Significant global investments are being made to overcome \gls{QKD} limitations and advance quantum technologies~\cite{qurecaQuantumInitiatives}. The US National Quantum Initiative~\cite{USQ}, EuroQCI under the EU Quantum Flagship~\cite{EuroQCI, qt}, and the UK’s Quantum Technologies Program~\cite{ukQ, UKQhub} are driving secure quantum communication development. Other major contributors include Germany, the Netherlands~\cite{quantumdeltaQuantumNetworks}, China, Japan, South Korea, India~\cite{dstIndia}, and Singapore~\cite{qurecaQuantumInitiatives}, fostering a competitive global push toward scalable quantum communication systems.
Beyond research, private companies such as ID Quantique~\cite{idquantiqueProducts}, Toshiba~\cite{toshibaToshibaQuantum}, and others~\cite{QKDCOMP, QKDCOMP2} offer \gls{QKD}-based solutions. The global quantum communication market, valued at USD 1.1 billion in 2023, is projected to reach USD 8.6 billion by 2032~\cite{quantum_communication_2024}

As \gls{QKD} gains traction as a secure communication technology, identifying relevant use cases is essential for maximizing its potential, driving adoption, and guiding further research. Various surveys, white papers, and commercial reports offer insights into practical implementations. Early studies, such as~\cite{distribution2010gs}, identified QKD use cases including off-site backup, enterprise MANs, critical infrastructure security, backbone protection, and high-security networks, detailing their operational aspects and challenges. A more recent survey by~\cite{gramegna2023fgqt} expands on these, highlighting emerging applications in metropolitan networks, healthcare, smart cities, and industrial automation, often integrating \gls{PQC} for enhanced security.

Several companies have demonstrated real-world QKD deployments. Toshiba~\cite{TOSIBAusecases} showcased use cases in genome data security, back-office protection, and secure transfers. ID Quantique applied QKD in banking, finance, government, and critical infrastructure sectors~\cite{idquantiqueApplications}. QNu Labs~\cite{qnulabsuse} explored enterprise and critical infrastructure applications, while QuantumCTek~\cite{quantuminfoTypicalCases_QuantumCTekx2013} focused on securing governmental communications. Quantum Xchange highlighted high-level applications of its QKD solutions~\cite{quantumxc}. These implementations illustrate QKD’s adaptability across diverse domains.

\heading{Motivation.} While numerous \gls{QKD} use cases exist, the currently available literature (including manufacturers’ resources and project reports) do not provide a comprehensive practical security analysis.
Specifically, sources in the literature often fail to critically examine (or even frequently omit) key aspects we view as crucial. 
This encompasses aspects such as security requirements, the concrete usage of the generated keys, which kind of data they aim to protect (and for how long), the protocol with which the proposed solution operates, network topology, and the concrete targeted security guarantees.
Further relevant aspects that are missing in such documents are whether the proposed solution 
involves additional cryptographic components (and the impact of potential failures of these components), and a discussion of/comparison with potential alternative approaches (such as pre-shared keys and \gls{PQC}).

\heading{Our contribution.} This paper addresses this gap by conducting a comprehensive, detailed and practical security analysis of available \gls{QKD}-based use cases to evaluate their feasibility and effectiveness. 
We evaluated use cases deployed in optical fiber networks, focusing on those with sufficient publicly available information to enable detailed evaluation. We systematically selected use cases from the literature, sorting them chronologically from older to more recent examples.
We critically analyze each use case based on parameters such as target sector, \gls{QKD} system employed (including technical details provided by \gls{QKD} providers and underlying technology), security goals, and type of data to be protected. We excluded use cases that lack adequate information for the analysis but listed them in the appendix for reference. Our analysis identifies strengths and gaps in the reviewed use cases, thus contributing to a general assessment of their practicality and effectiveness. Furthermore, we provide recommendations on how to improve the solutions discussed per use case and explore whether similar outcomes could be achieved using classical cryptographic systems, thus offering a comprehensive perspective on the applicability of \gls{QKD} solutions.

\heading{Organization of this paper.} We begin by providing necessary background in \cref{background} and details on the use cases to be analyzed in \cref{use-cases}. We then describe our approach to use case analysis in \cref{analysis-method}. In \cref{analysis}, we apply this method to analyze use cases, critically comment on the use cases and offer recommendations for improvement. Finally, in \cref{conclusion}, we conclude the paper.

\printglossary[type=\acronymtype,style=list,nonumberlist]

\section{Background}\label{background}
	
\gls{QKD} is a comparatively new technology, that allows two parties to establish a shared key. In this section, we will give an explanation of the terms used and a brief overview of QKD including trade-offs, alternatives such as \gls{PQC} and pre-shared keys, assumptions, and related schemes.

\subsection{Security properties}

A cryptographic solution can provide different security properties. Our main focus in this work is on confidentiality and authenticity as the most commonly required ones, but we note that other properties such as anonymity may be as, or even more important, depending on the use case.

\subsubsection{Confidentiality}
Confidentiality intuitively means that the content of an interaction remains hidden from everyone but the involved parties.
As far as encryption of large plaintexts is concerned, this means that no outside party should be able to learn any information about it, except for its length. In case of QKD and key-exchange protocols, an equivalent but usually preferred definition is that the key is indistinguishable from a random key sampled from the same distribution.
This property is inherently vulnerable to \glsxtrshort{HNDL} attacks, and security parameters are generally chosen to make decryption infeasible, even for extended periods.

Some protocols achieve confidentiality by encrypting messages to keys that are in use for a long time.
If such a key is corrupted at some point, this allows the decryption of all future and past communication.
To deal with this threat, protocols will often implement stronger versions of general confidentiality, that require graceful treatment of this scenario: Forward Secrecy and \glsxtrfull{PCS}.

\textit{Forward Secrecy}, also occasionally known as Pre-Compromise Secrecy or, misleadingly, as Perfect Forward Secrecy (there is nothing perfect about it) is the property that a protocol prevents the decryption of old messages, even if a key is corrupted at some point. To achieve forward secrecy, a protocol can update its session key at regular intervals and remove old ones, for example with a new key exchange such as \gls{QKD} or with a symmetric ratchet.

\textit{\glsxtrfull{PCS}}, also known as Backward Secrecy, on the other hand is the property that a protocol can recover from a corruption and get back to a point of full confidentiality, even if a party’s state gets corrupted at some point~\cite{ccg16}. To achieve \glsxtrlong{PCS}, a protocol can update its session key with fresh randomness at regular intervals with a new key exchange, for example with \gls{QKD} or an \glsxtrfull{AKE}.

\subsubsection{Authenticity}

Authenticity intuitively means that the parties participating in an interaction are indeed the parties that they claim to be and that their messages contain the content that they actually transmitted. Sometimes the latter part is treated as the distinct property `integrity', though we will use the term authenticity to also cover integrity.

\subsection{Computational and information-theoretical security}
Cryptographic security notions can offer resistance against different classes of attackers.
The most powerful notion in this respect is \emph{perfect} security, which is defined as unconditional security against computationally unbounded adversaries in all cases.
A slightly weaker version of this is \emph{statistical} security, which allows an adversary to break security in a very small and random number of cases, but without any influence on whether he \emph{gets lucky}.
Both of these can be summarized as information-theoretical security, since they can be proven directly from information theory. However, information-theoretical security is sometimes also used to refer to perfect security only.

The more common alternative is computational security, in which computationally bounded adversaries should only have a negligible chance of breaking a scheme. This class can be further subdivided based on whether the adversary has access to a quantum computer or not: In the former case we call the adversary a \emph{quantum adversary} and in the latter a \emph{classical adversary}.

In the context of QKD, we gain one further class: Schemes whose security can be proven with quantum information theory, which adds quantum mechanical assumptions to information theory. We call notions that fall into this category \emph{quantum-information-theoretically} secure.

\subsection{One-Time Pads and Message Authentication Codes}

\glsxtrfullpl{OTP} are a method in which a plaintext is \glsxtrshort{XOR}ed with a random key of equal length. The key is used only once before being discarded.
OTPs at times are not even viewed as an encryption scheme in the stricter sense due to the inability to reuse keys, and since the length requirement on the key poses severe limitations.
On the other hand,  correctly used OTPs are the only way to achieve unconditional confidentiality that requires no further conditions or assumptions.
A scheme with this level of security could be called \emph{perfectly} confidential.

On their own, however, OTPs offer no authenticity -- it is trivial to manipulate ciphertexts with (partially) known plaintexts in a way such that they decrypt to other messages. To prevent this, we require an authenticity mechanism, which is usually achieved via message authentication.
The strongest possible way to achieve this is with \glsxtrfullpl{MAC} based on universal hash families, originally introduced by Carter and Wegman~\cite{carterwegman, wegmancarter}.
\Glspl{MAC} based on universal hashing still can be attacked with a very small chance -- the very small chance is that an attacker correctly guesses the authentication tag due to sheer luck,
but there exist no successful attack strategies beyond guessing.
While such schemes thus aren't perfectly secure, they offer \emph{statistical} authenticity.

\subsection{Encryption with pre-shared keys}\label{background:OTPwithPSK}
Considering the significant cost of setting up quantum channels between two parties, it is worthwhile to investigate whether the cost of doing this exceeds the cost of simply exchanging the necessary key material directly (in person) between the endpoints.
Although such an approach does not scale to universal use on the Internet, it can be viable with a small number of well-known endpoints and predetermined connectivity. Some of the use cases that we analyzed fall into that category, even when envisioned as fully deployed.

We distinguish two scenarios, depending on the encryption method.
First, the \gls{OTP} can be combined with a statistically secure \gls{MAC} to achieve \gls{ITS}. Since \glsxtrshort{OTP} requires a key that is at least as long as the data that is being transmitted, this requires an initial trusted offline distribution of a large number of pre-shared key bits, each of which has to be discarded after use. For example, this distribution could be done by a trusted courier. Moreover if more data needs to be transmitted, then additional trusted offline key distributions are required. The key material for an \glsxtrshort{OTP} must be deleted after use, both to prevent accidental reuse, but also to reduce the attack surface by preventing the decryption of past ciphertexts as a consequence of a later compromise of the keys (`forward secrecy'). Once this is done on both ends, this method guarantees everlasting confidentiality.
A second option is to use computationally secure symmetric cryptography, for example \gls{AES} with \gls{GMAC}, so that only a small number of pre-shared key bits are required. Only an initial offline distribution is required: as estimated below, one could simply distribute enough key bits initially to last for the effective lifetime of the system. Alternatively, a single small key can be distributed initially, which is stretched via a symmetric ratchet (the ratchet even be based on AES to minimize the computational assumptions in the system). This ratchet would also provide forward secrecy in case of key leakage via (for example) a device compromise.

Modern mass storage has become incredibly cheap, to the point where micro SD cards with a capacity of 1 \glsxtrshort{TB} are commercially available for under 100€.
At the rate of consuming one 256-bit key per minute for AES, this would be enough capacity to store key-material for almost 60.000 years\footnote{1 \glsxtrshort{TB} SD cards can store enough 256 bit keys for $10^{12} \cdot 8/256/(60\cdot 24\cdot 365.25) = 59415$ years, when one key is used every hour.} at the cost of exchanging a physical item once, requiring no assumptions besides the availability of a good source of randomness and the correct identification of the peer when handing over the key-material. We compare the usage of pre-shared keys with QKD in \cref{tab:security_comparison}.

The downside of pre-shared keys compared to QKD and AKE is that this method does not provide \gls{PCS}. \gls{PCS} is not achieved because key-material cannot easily be generated on the fly. This is problematic after a breach that may have corrupted key-material intended for later use, because this could result in the unavailability of key-material until new key-material is exchanged. We remark though that even QKD and AKE could run into issues in that scenario, as this kind of breach could also affect key-material for authentication. Whether the cost of the new individual exchange is cheap or expensive heavily depends on the use case in question.

\subsection{Asymmetric cryptography, PQC}\label{background:asymmetric}

Traditionally, the problem of establishing a shared secret key between two parties is solved using asymmetric cryptography. Initially, this came mainly in the form of asymmetric encryption based on the \glsxtrfull{RSA} algorithm, a public-key cryptosystem that relies on the computational difficulty of factoring large composite numbers~\cite{rsa_1978}. At this point, we also see a large number of protocols based on versions of the \glsxtrfull{DHKX}~\cite{dh_1976}. We compare the usage of an authenticated key exchange using asymmetric cryptography with QKD in \cref{tab:security_comparison}.
These techniques scale exceptionally well and by now protect most of the communications infrastructure upon which the World Wide Web operates. Additionally, they also protects much of the Internet and most digital communication.
On the other hand, like all asymmetric cryptography, RSA and DHKX rely on computational assumptions.
Shor's algorithm additionally renders them vulnerable to quantum attacks.

\gls{PQC} is a name for classical algorithms that are expected to withstand quantum attacks.
The first asymmetric PQC solutions have been standardized by \glsxtrshort{NIST},
encompassing ML-KEM~\cite{MLKEM} (based on Kyber~\cite{Kyber}) and \gls{HQC}~\cite{NISTPQC-R4:HQC22}  (FIPS to appear) for key establishment
as well as several digital signing algorithms,
called ML-DSA~\cite{MLDSA} (based on Dilithium~\cite{Dilithium}),
SLH-DSA~\cite{SLHDSA} (based on SPHINCS+~\cite{SPHINCS}), and FALCON~\cite{Falcon} (FIPS to appear). Additionally, some European government agencies have approved the use of the conservative KEMs
Classic McEliece~\cite{NISTPQC-R4:ClassicMcEliece22} and Frodo~\cite{NISTPQC-R3:FrodoKEM20}.
Lastly the \glsxtrshort{IRTF} standardized the two stateful signature schemes XMSS~\cite{rfc8391} and LMS~\cite{RFC:LMS}.
We provide an overview of all of these in  \cref{tab:StandardizedPQC}.

\subsubsection{Strengths} Here we list some benefits of using   asymmetric cryptography including PQC.
\begin{enumerate}
    \item \textbf{Integration with classical systems:} The fact that these algorithms can still be run on classical machines has one advantage:
at least in some use cases, the infrastructures that so far used RSA/DHKX can accommodate them with reasonably work (e.g., without switching to a quantum channel, and without adapting the bigger protocols surrounding the algorithm in a major way). For example, ML-KEM/Kyber have already been deployed in Chrome and Firefox to establish \gls{TLS} connections~\cite{PQCinChrome}, by Amazon for its AWS Key Management Service~\cite{PQCinAmazon},
and by iMessage~\cite{PQCinIMessage} and Signal~\cite{PQCinSignal} (whose key agreement mechanism PQXDH is also used by WhatsApp).
\item \textbf{Authentication}: Asymmetric cryptography is able to provide one-sided authenticity, which is, for example, the typical need in the World Wide Web and similarly desirable for sending anonymous messages. This property is much harder to achieve with QKD (without relying on asymmetric cryptography).
\end{enumerate}
\subsubsection{Challenges} Here we list some major challenges.
\begin{enumerate}
    \item A caveat with regards to integrating post-quantum algorithms into protocols is that the algorithms tend to have a larger data footprint and/or are slower to compute than pre-quantum ones, thus potentially requiring some protocol adjustments.
These algorithms cannot be viewed as general drop-in replacements without considering the concrete use case at hand - e.g., the sizes of ML-KEM's public keys are significantly larger than those of RSA/ECC, which in turn significantly increases data transmission in certain use cases and might prohibit usage of ML-KEM in low-resource systems.
\item Compatibility issues can arise if some of the systems have been updated and other related ones have not. (E.g., services using the updated TLS protocol can encounter the situation that the ClientHello message does no longer fit into a single TCP packet due to the bigger key size and is expanded into multiple packets, which is not digestible for services who have not performed the respective update.)
\item A common misconception on the approach towards the implementation of PQC is that this is purely a software issue. However, for hardware-efficient implementation, these algorithms or underlying primitives are more likely to be implemented on \glsxtrfullpl{FPGA} or on \glsxtrfullpl{ASIC}. Only such implementations will provide the necessary hardware acceleration to support the migration of applications of large scale to PQC. If any vulnerabilities are however exposed, it becomes difficult to update the affected hardware logic, to the point where on-site intervention or even hardware replacements may be required.
\end{enumerate}

\subsubsection{Assumptions of classical cryptography (including PQC)}
Here we list all the assumptions for classical cryptography which also includes PQC.

\begin{enumerate}
    \item \textbf{Computational assumptions}: 
With PQC having been standardized and the first algorithms having been deployed, we note that the main caveat of PQC is that of most classical cryptography:
all computationally secure primitives have to rely on computational hardness assumptions.
The only principal difference between PQC algorithms and their predecessors lies in the concrete computational assumptions on which the algorithms rely. It should be noted that the PQC assumptions involve mathematical principles that are often more complex than the assumptions underlying pre-quantum cryptography, such as the hardness of factoring integers or computing discrete logarithms.

When it comes to schemes that are standardized, approved, or chosen for standardization by major national or international standardization bodies, we find three general categories of assumptions:
\begin{enumerate}
    \item \textbf{Assumptions on the difficulty of \gls{LWE}:} \gls{ML-KEM} and \gls{ML-DSA} use an assumption called Module-\gls{LWE}~\cite{langlois_worst-case_2015}, Falcon uses an assumption called \gls{NTRU}~\cite{goos_ntrusign_2003}, 
   and Frodo relies on (not module) \gls{LWE}~\cite{regev_lattices_2009}.
The first two of these lattice assumptions are comparably `young' -- while the scientific community has been conducting dedicated cryptanalysis for several years, the assumptions have not encountered cryptanalytical attention for the same period of time as the ones underlying \gls{RSA} and \gls{DHKX}.
\item \textbf{Assumptions on the difficulty of hash functions: }
\gls{SLH-DSA}, \gls{XMSS}, and \gls{LMS} are highly conservative designs that only rely on assumptions about hash functions.
The primary drawback of \gls{SLH-DSA} are its large signatures and slow runtime performance, whereas \gls{XMSS} and \gls{LMS} suffer from the fact that they are stateful signature schemes, which require a signer to evolve a state for every signature.
\item \textbf{Assumptions on the difficulty of decoding errors:}
Classic McElice and \gls{HQC} rely on the difficulty of decoding errors. While the general problem has long been known to be NP-hard~\cite{berlekamp_inherent_1978}, this does not inherently translate into specific instantiations being as difficult to solve. Classic McEliece is a slight variant of a scheme that was designed in 1978, making it almost as old as \gls{RSA}, and has since stood secure. While it offers great performance for encapsulation, decapsulation, and ciphertext-size, it requires large public keys (starting at 261~kB at NIST Level~1) that make it suboptimal for practical use. While \glsxtrshort{NIST} did not chose it, using it was approved by \glsxtrshort{BSI}, \glsxtrshort{ANSSI}, and \glsxtrshort{NCSC}.
\gls{HQC} on the other hand is a much younger scheme with a more balanced performance-profile relying on the difficulty of decoding random quasi-cyclic codes that \gls{NIST} chose as fallback for \gls{ML-KEM}.
\end{enumerate}

\item \textbf{Implementations assumptions}: 
Besides their specific underlying assumptions, all classical crypto (including PQC) can of course also suffer from bad implementations:
While implementations that produce different results than they should are not unheard of, but relatively easy to detect, a more common issue are vulnerabilities to side-channel attacks. Many of these fall into one of two categories:
\begin{enumerate}
    \item \textbf{Timing-attacks} can reveal information about branches, cache-accesses, and even the arguments to instructions of fundamental operations such as multiplication and division. Conservative implementations therefore avoid calling the affected instructions with values that could reveal secrets if they leaked and many modern algorithms are designed to avoid their use.
    \item \textbf{Power-analysis attacks} can work by measuring the precise power-consumption of a device while it performs cryptographic operations.
While they represent an extremely powerful class of attacks, they require access to precise information about energy-consumption, which is not necessarily easy to acquire. In particular it may be necessary to receive physical access over the attacked end-point.
As a consequence many software-implementations disregard this class of attacks as out-of-scope for their security-model. Hardware implementations however generally take this vector into consideration.
\end{enumerate}
Next to these primary categories, there are many other possible channels, such as electromagnetic radiation or even acoustic signals. Besides preventing leakage, there is also some research into designing algorithms that can survive some degree of leakage, similar to how QKD solutions attempt to deal with the problem.

\end{enumerate}
\begin{table}
\caption{List of asymmetric post-quantum schemes that are standardized (\textbf{S.}), approved (\textbf{A.}), or chosen for standardization (\textbf{C.}). In cases there were multiple options at a given security-level (\textbf{L.}), we give the sizes of the smallest one.}
\label{tab:StandardizedPQC}
\begin{tabular}{|>{\raggedright}p{2.2cm}|>{\raggedright}p{1.3cm}|>{\raggedright}p{1.4cm}|>{\raggedright}p{1.4cm}|>{\raggedright}p{3.2cm}|}
\hline
\textbf{Scheme} & \textbf{Type} & \textbf{Status} & \textbf{Based on} & \textbf{Sizes} [Bytes]\\
\hline
\hline
    \gls{ML-KEM} (“Kyber”)     & \gls{KEM}                & \textbf{S.} (\gls{NIST})             & Lattices (\gls{M-LWE})
    & L.~1: pk:~800, c:~768  L.~3: pk:~1184, c:~1088  L.~5: pk:~1568, c:~1568 \\ \hline
    Frodo                & \gls{KEM}                & \textbf{A.} (\gls{BSI}, \gls{ANSSI}, \gls{NCSC}) & Lattices (\gls{LWE})
    & L.~1: pk:~9616, c:~9720 L.~3: pk:~15632, c:~15744  L.~5: pk:~21520, c:~21632 \\ \hline
    Classic McEliece     & \gls{KEM}                & \textbf{A.}  (\gls{BSI}, \gls{ANSSI}, \gls{NCSC}) & Codes
    & L.~1: pk:~261120, c:~96 L.~3: pk:~524160, c:~156  L.~5: pk:~1044992, c:~208 \\ \hline
    \gls{HQC}                  & \gls{KEM}                & \textbf{C.} (\gls{NIST})             & Codes
    & L.~1: pk:~2249, c:~4497 L.~3: pk:~4522, c:~9042  L.~5: pk:~7245, c:~14485 \\ \hline
    \hline
    \gls{ML-DSA} (“Dilithium”) & Signature          & \textbf{S.} (\gls{NIST})             & Lattices (\gls{M-LWE})
    & L.~1: pk:~1312, sig:~2420 L.~3: pk:~1952, sig:~3293 L.~5: pk:~2592, sig:~4595 \\ \hline
    \gls{SLH-DSA} (“SPHINCS+”) & Signature          & \textbf{S.} (\gls{NIST})             & Hashes
    & L.~1: pk:~32, sig:~7856 L.~3: pk:~48, sig:~16224 L.~5: pk:~64, sig:~29792    \\ \hline
    \gls{FN-DSA} (“Falcon”)    & Signature          & \textbf{C.} (\gls{NIST})             & Lattices (NTRU)
    & L.~1: pk:~1281, sig:~666 L.~5: pk:~1793, sig:~1280 \\ \hline
    \hline
    \gls{XMSS} (RFC~8391)      & Stateful Signature & \textbf{S.} (\gls{IRTF})             & Hashes
    & L.~5: pk:~64, sig:~2500   \\ \hline
    \gls{LMS} (RFC~8554)       & Stateful Signature & \textbf{S.} (\gls{IRTF})             & Hashes
    & L.~5: pk:~56, sig:~2664   \\ \hline
\end{tabular}
\end{table}

\subsection{Quantum Key Distribution}

\Gls{QKD} is a mechanism that uses a quantum channel, a channel that allows the transmission of quantum states (generally speaking qubits), to establish a shared secret between the endpoints.
On its own, this mechanism would be inherently vulnerable to man-in-the-middle (MitM) attacks. 
It is thus necessary to perform the procedure in an authenticated way, by means of an authenticated channel (which may be classical).
Usually, `\gls{QKD}' refers to the combined mechanism and encompasses the authenticated channel.
The resulting shared secret is said to offer confidentiality based on the postulates of quantum mechanics, assuming that the authenticated channel is indeed information-theoretically secure and that executions of the protocol do not deviate from its abstract design.
This is often claimed to offer \glsxtrfull{ITS}, though we would argue that this can be confusing -- security is based on the additional assumptions that underlie QKD, such as the postulates of quantum mechanics.
We expand on the assumptions underlying QKD in~\cref{background:QKDassumptions}.
To make a clear distinction between information-theoretic security and its extension with quantum physics postulates,
we will use the term \textbf{\glsxtrfull{Q-ITS}}.

The keys established by QKD can be used with a \glsxtrshort{OTP} and a statistical authentication mechanism to achieve quantum-information-theoretic confidentiality and authenticity, but this requires a large amount of key material.
Many real-world schemes will thus instead combine QKD with classical, computationally secure symmetric encryption mechanisms, resulting in computational security with additional quantum assumptions.

\subsubsection{Assumptions about QKD}\label{background:QKDassumptions}

Although QKD-based solutions are often advertised as solutions that dispense with the computational assumptions underlying public-key cryptography, it is important to note that they come with their own assumptions and limitations. 
Most of these assumptions
are protocol-dependent, but below are some fundamental ones independent of this choice of the QKD protocol~\cite{renner_quantum_2022}.

\begin{enumerate}
    \item \textbf{Quantum theory is correct:} We assume that quantum theory accurately predicts measurement outcomes, as numerous experiments have confirmed. However, while this assumption is sufficient for quantum cryptography's security, it may be stronger than necessary. The security relies only on key elements of quantum theory such as state space structure, operations on it, and the prohibition of superluminal communication. These principles show that quantum states cannot be cloned, and entanglement is monogamous. Other aspects, like the Schrödinger equation, are not essential. Even if quantum theory is adjusted, as it likely will be to incorporate gravity, as long as these core principles remain approximately valid, quantum cryptography remains secure.
    \item \textbf{Quantum theory is complete:} Completeness means there is no extended theory that can make better predictions used for the security analysis of QKD. 
    It has been shown that the completeness of quantum theory follows from its correctness and the existence of free randomness~\cite{colbeck_no_2010}. In QKD, security is guaranteed against any attack within quantum theory's framework. \textit{Completeness ensures that an adversary cannot gain more information about the key than what quantum theory predicts}.
    \item \textbf{Devices do not leak any relevant/useful/secret information:} It is assumed that devices like single photon detectors, the QRNGs, and the classical computer only leak information as specified in the protocol. For instance, the raw key stored on the classical computer must not be leaked externally. To ensure this, the hardware is typically required to be properly shielded. Whether this is fundamentally possible, remains an open question~\cite{bernsteinQKDvsPhysics,renes_are_2020}.  It can be noted that for certain components involved such as phase modulators, intensity modulators, lasers, and their associated control electronics, security proofs have been developed that account for possible information leakage. These works demonstrate that even when some level of side-channel leakage is present, it is still possible to establish security under well-defined assumptions. This highlights the progress in security models that can tolerate device imperfections while maintaining rigorous security guarantees. Understanding and quantifying the impact of such leaks remains an active area of research, particularly in practical implementations of QKD~\cite{curras2023security,pereira2020quantum,tamaki2014loss,gottesman2004security}. 
    \item \textbf{Existence of Authenticated Classical Channel:} Any quantum key distribution protocol relies on the existence of an authenticated classical channel~\cite{paterson_quantum_2004}. 
    
    Common options for this are the use of ITS MACs with a pre-shared trusted key and the use of classical cryptographic signatures.
    The former option has the downside that establishing that pre-shared secret is still a hard problem, and that both parties can be impersonated to each other in case a key becomes compromised, whereas the latter ties the authenticity of the scheme to computational assumptions.

    It is possible to establish the authenticity with signatures once and then switch to ITS MACs, though this still maintains the issues with regards to authenticity after the compromise of a key and still requires computational assumptions, though now with a potentially smaller attack-window.
    
    \item \textbf{The protocol is implemented correctly:} In addition to fundamental assumptions about the underlying physical theory, various assumptions can be made about the protocol's implementation. The more assumptions we make, the easier the security proof becomes, as each assumption limits the range of possible attacks by an eavesdropper. For example, assuming all detectors have the same detection efficiency simplifies the security proof by excluding that threat. However, the proof is only valid if the assumptions hold—any deviation in implementation can compromise the protocol's security. 
    \item \textbf{Use of Quantum repeater or Trusted nodes for long-distance communication:} From the implementation point of view, QKD faces limitations in long-distance communication as quantum signals attenuate significantly over long distances, and increasing this range without compromising security is challenging. However, QKD works effectively over short distances because the attenuation and noise levels are manageable, ensuring high security and low error rates. For long-distance communication, either quantum repeaters or trusted nodes are necessary. Quantum repeaters use principles like entanglement swapping and quantum memory to extend the communication range, but they are not yet fully developed for large-scale practical use. Trusted nodes are more practical with current technology but introduce vulnerabilities if the nodes are compromised. Thus this assumption is not inherent but rather a technological assumption.

\end{enumerate}

\subsection{Types of QKD service}

\begin{figure}
\centering
\includegraphics[width=0.9\textwidth]{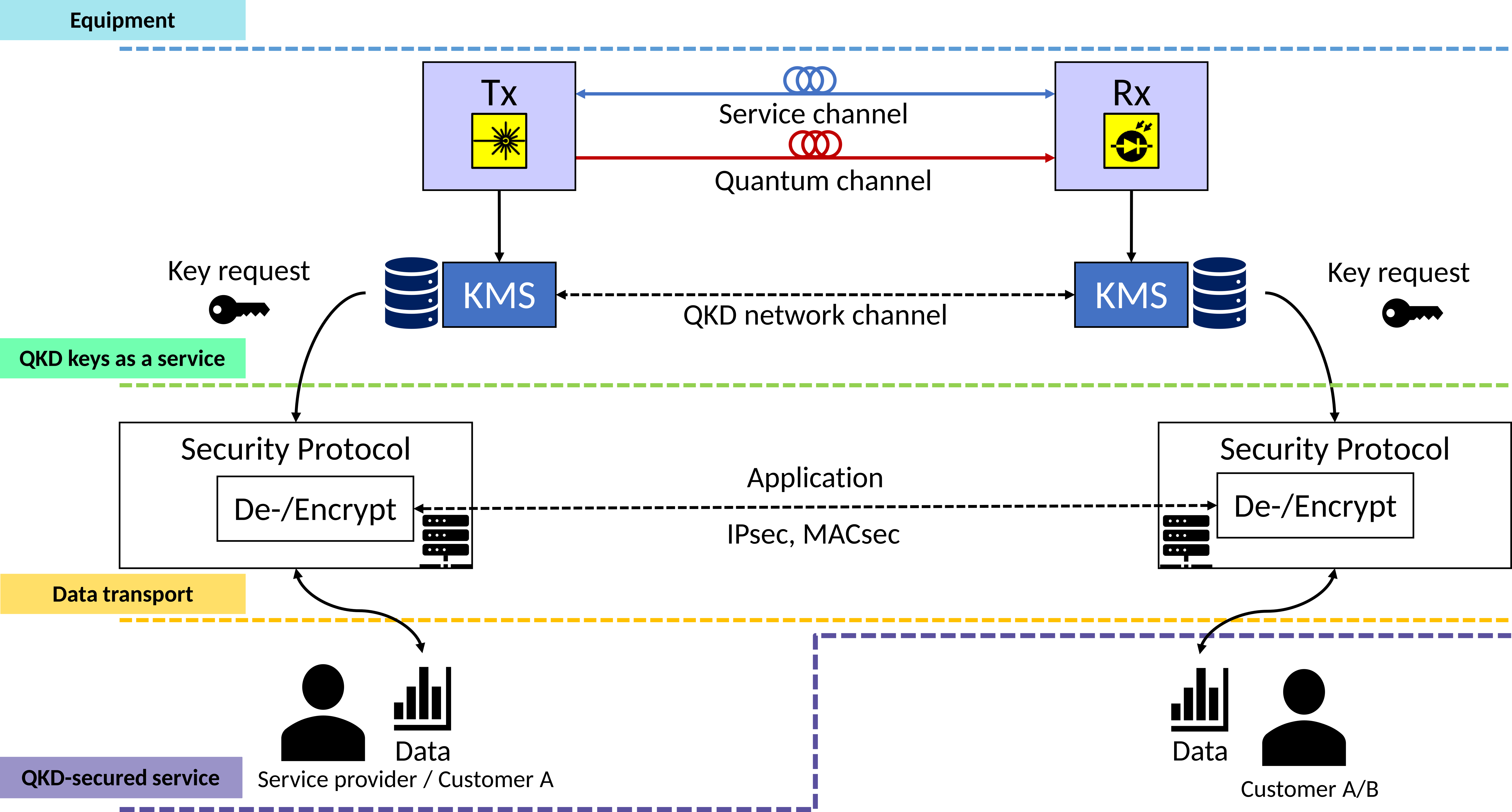}
\caption{Overview of systems involved in the encryption of data using QKD keys. Different types of service provide different functionality and determines which systems the user is required to operate.}\label{fig-cuts}
\end{figure}

There are multiple ways in which a user can use QKD. Fig.~\ref{fig-cuts} shows which ones we encountered in our use case analysis and shows what is operated by the end-user and what is operated by the service provider. Horizontal dotted lines with a color have an accompanying service offering written in a box on the left. All components above the respective line are operated by the service provider, while the components below the respective line are operated by the end-user. Vertical communication lines are within a node and are assumed to be physically secured, while horizontal communication lines can be accessed by an attacker. 

\begin{itemize}
    \item Equipment. An end-user can operate all the necessary equipment himself, including the QKD \glsxtrfull{Tx} and \glsxtrfull{Rx}.
    \item QKD keys as a service. In QKD keys as a service, the QKD equipment is operated by a network operator and the QKD equipment can be shared with multiple customers. The customer might pay a subscription fee or pay per QKD key. The provider operates the \glsxtrfull{KMS} and allows the customer to connect to the \glsxtrshort{KMS} to retrieve his keys.
    \item Data transport. The customer does not use the QKD key himself but the key is used by the provider to securely transport the data of the customer. The provider is responsible for the encryption and decryption of the user data. The provider might for example offer encrypted transport based on \glsxtrshort{MACsec} or \glsxtrshort{IPsec} with QKD keys as a service.
    \item QKD-secured service. The provider provides a service to the customer in which QKD is used and in which one endpoint of the QKD system is used by the service provider and the other endpoint by the customer.
\end{itemize}

\subsection{QKD network topology}

\begin{figure}[h]
\centering
\includegraphics[width=0.9\textwidth]{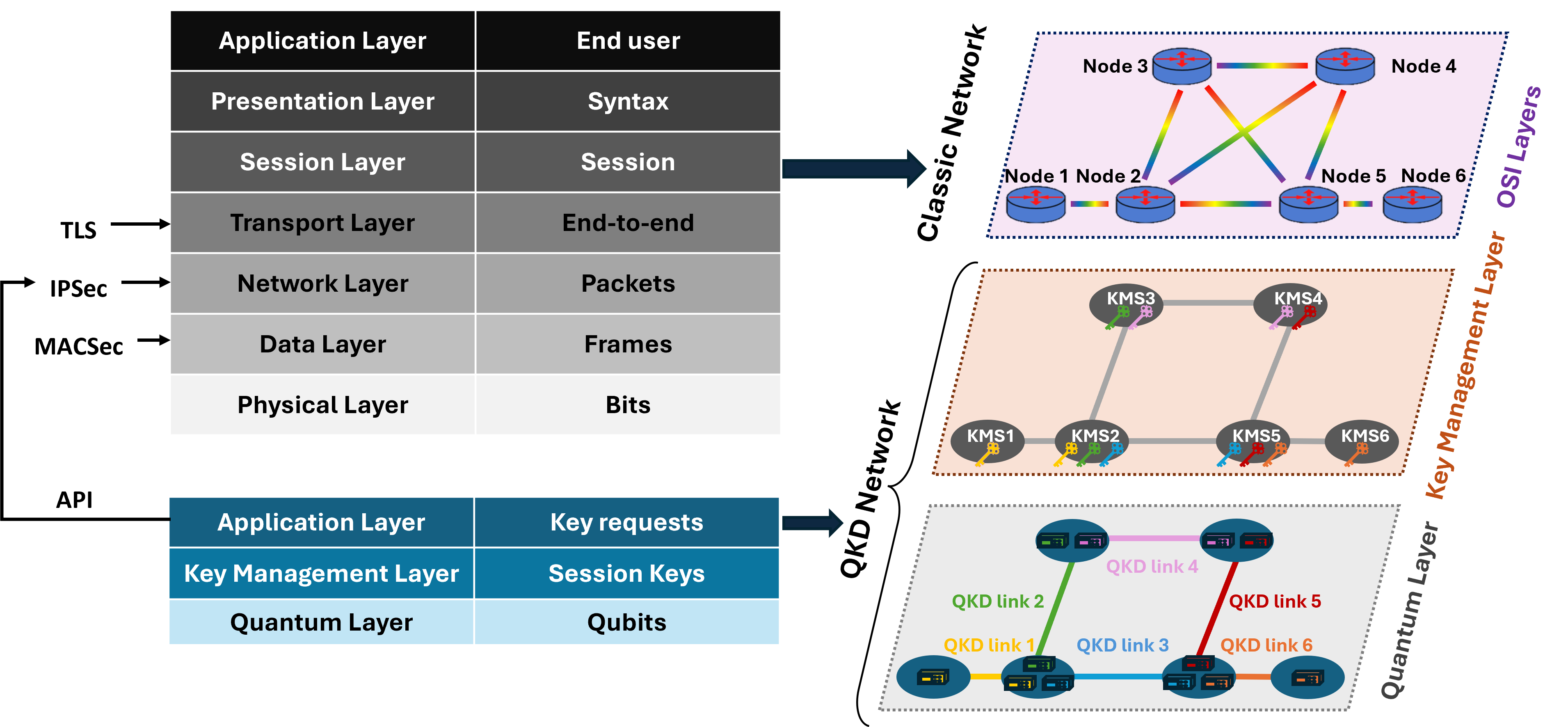}
\caption{General architecture of an optical network supporting QKD. Protocol stacks of classic optical network (top-left) and QKD network (bottom-left).}\label{fig-net}
\end{figure}

The purpose of a key exchange is that only two parties share the key material, which perfectly fits the basic \glsxtrfull{P2P} connections. However, when dealing with more complex network topologies, a device-based technology such as QKD requires a more complex architecture (see  Fig.~\ref{fig-net}) ~\cite{ITUTY3810}. Because deploying fibers is an expensive and time-consuming operation, any fiber operator would prefer sharing their pre-existing fiber infrastructure over deploying an entire new one only for QKD. A typical architecture of shared infrastructure includes two parallel optical networks~\cite{Sasaki2011Tokyoqkd, Rubio_secure6G}: the classical network, which is based on the OSI model of 7 layers, and the QKD network, which includes two subnetworks referred as \glsxtrfull{KM} and quantum layers. The connection between the two classical and QKD networks is provided by an application layer that manages keys requests, which are managed by the QKD network. As the name suggests, the KM layer connects the \gls{KMS}s assigned to each node. The KM and application layers can easily share the infrastructure with standard \glsxtrfull{WDM}. However, this is not the case for the quantum layer, which connects QKD \glsxtrshort{Tx} and \glsxtrshort{Rx}, because of the quantum channel. The sharing of fibers in classical networks presents numerous technical challenges, such as crosstalk, scattering, and additional filters so that quantum channels can bypass optical amplifiers~\cite{Mehdizadeh2024qkdcoex}.

QKD protocols are executed in the quantum layer. Typically, service channels must be synchronized with quantum ones, thus sharing the same or parallel fibers is preferred. However, no standardization has been proposed yet on the QKD protocol to adopt, making it difficult for fiber operators to integrate QKD into their infrastructure~\cite{Saez2024standard}. There are two main families of QKD protocols: 
\glsxtrfull{DV-QKD} and \glsxtrfull{CV-QKD}. \Cref{tab:QKDprotocols} reports some of the QKD protocols that are closest to being deployed on real networks.

\begin{table}[h!]
\centering
\caption{Summary of most common QKD Protocols according to their maturity level. Achieved maximum distance with related SKR are reported.}
\label{tab:QKDprotocols}
\begin{tabular}{|>{\raggedright}p{2cm}|>{\raggedright}p{2cm}|>{\raggedright}p{1.3cm}|>{\raggedright}p{2.2cm}|>
{\raggedright}p{2cm}|>{\raggedright}p{2.5cm}|}
\hline
\textbf{Protocol} & \textbf{Quantum state} & \textbf{Cost} & \textbf{Maximum distance [km]} & \textbf{SKR [Mb/s]} & \textbf{Maturity level} \\
\hline
    \textbf{BB84 (DV) ~\cite{toshiba_products} }
    & Single photon
    & medium
    & 150 
    & 0.3 @50 km
    & Variants of BB84 available as commercial products. \\
\hline
    \textbf{COW (DV) ~\cite{IDQ_products} }
    & Coherent pulses
    & medium
    & 90 
    & 0.05
    & Available as commerical product. \\
\hline
    \textbf{MDI (DV) ~\cite{liu_experimental_2023} }
    & Pair of photons
    & high
    & 442 
    & 2.19 $10^{-4}$
    & Deployed on real networks. \\
\hline
    \textbf{TF (DV) ~\cite{Chen2021TFQKD} }
    & Pair of photons
    & high
    & 511 
    & 3.37 $10^{-9}$
    & Deployed on real networks. \\
\hline
\textbf{Gaussian (CV) ~\cite{GM_CVQKD_2016, CVQKD_GM_trial}}
    & Coherent state
    & low
    & 20 
    & 0.03
    &  Field demonstrations.\\
\hline
\textbf{QAM (CV) ~\cite{CVQKD_QAM_trial}}
    & Coherent state
    & low
    & 16 
    & 35
    &  Field demonstration.\\
\hline
    \textbf{QPSK (CV) ~\cite{Liao2020CVQDKDM} }
    & Coherent state
    & low
    & 20 
    & 10
    &  Deployable system under development.\\
\hline
\end{tabular}

\end{table}

We can observe that all proposed protocols are limited in \glsxtrfull{SKR} when compared to typical data rates on fiber communications, which are on the order of hundreds of gigabits. This limitation prevents the use in real time of QKD for ITS symmetric encryption schemes that require keys comparable in size to the data, such as OTP. SKRs are still sufficient for symmetric encryption algorithms with smaller keys such as AES, although this comes with a cost in terms of the overall security of the system.
Despite sharing the same fiber infrastructure, classical data transmission and QKD are virtually separated into classical and quantum networks. One of the main motivations behind this separation is the incompatibility of QKD with some classic photonic components such as amplifiers. Other motivations are the difference of multiple orders of magnitude between SKR and standard data rates, and between the power of quantum and classic signals. Moreover, the distance limitation of QKD combined with the absence of quantum repeaters, makes the presence of chains of intermediate trusted nodes necessary~\cite{Quoc2008trustednode}.\\ 
DV-QKD protocols were the first to be formulated and demonstrated, thus they are currently the most mature technologically and are already available as commercial products as shown in \cref{tab:QKDprotocols}. However, DV-QKD protocols have significant technological limitations and challenges that limit their deployment in realistic scenarios. The first limitation is the cost, mainly due to the single-photon detectors, which are expensive and are not standard telecom devices. The second limitation is the high sensitivity to crosstalk due to the extremely low power of the quantum channel compared to classic ones. There are, however, efforts to overcome this limitation~\cite{gavignet_co-propagation_2023}. A promising family of protocols are \glsxtrfull{MDI-QKD} protocols, which rely on an intermediate node for measuring two different streams of qubits. The main advantage of MDI-QKD is that the security does not depend on the implementation of the receiver, although the key rates are significantly lower compared to BB84~\cite{liu_experimental_2023}. An additional benefit of MDI-QKD is the longer distances compared to other DV-QDK protocols. Another family of measurement independent protocols is Twin-Field (TF)-QKD, which enables to increase the distance between two nodes even further~\cite{TFQKD}. TF-QKD has been reported to have achieved more than 500 km on field deployment ~\cite{Chen2021TFQKD}, although the SKR was significantly lower. The maturity of systems based on measurement-independent protocols is still limited to network deployments, which is promising but still behind compared to other DV-QKD protocols in \cref{tab:QKDprotocols}.  \\
CV-QKD protocols have the advantage of using coherent receivers with detectors more similar to those used for classical optical communications. This advantage enables not only the reduction of costs, but also the compatibility with equipment from telecom operators and WDM systems~\cite{Ware2022WDMCVQKD}. Another advantage is that \gls{CV-QKD} is more resilient to classical noise~\cite{hajomer_coexistence_2025}. CV-QKD protocols can be based on \glsxtrfull{GM} or \glsxtrfull{DM} such as \glsxtrfull{QAM} and \glsxtrfull{QPSK}~\cite{Liao2020CVQDKDM}. DM based protocols show some advantages compared to GM: higher reconciliation efficiency of practical error correction schemes, simpler implementation, and higher capacity in the very low \glsxtrfull{SNR} regime, which is a very common regime for CV-QKD schemes~\cite{Djordjevic2019CVQKD}. These properties make DM CV-QKD protocols optimal and scalable options for deployment into classical optical networks. However, the signal-to-noise ratio in CV-QKD protocols rapidly decreases with channel losses, thus limiting the use to shorter distances, as shown in ~\cref{tab:QKDprotocols}. In terms of maturity, CV-QKD relies on standard photonic components for telecommunications, which make it easier to demonstrate and deploy \cite{CVQKD_networkintegration}. However, significant effort is required in terms of signal post-processing and calibration. GM and QAM-DM CV-QKD system have been demonstrated in a field demonstrations \cite{GM_CVQKD_2016, CVQKD_QAM_trial} over deployed fiber links below 20 km. DM CV-QKD systems have been shown to achieve key rates beyond 1 Mbps in short distances, although significant effort is still needed for deployment in a real optical network \cite{CVQKD_progress, CVQKD_HighSKR, CVQKD_Bob_demo}. Compared to DV-QKD a significant limitation of CV-QKD protocols is their security that is based the assumption of perfect states separation, which cannot be guaranteed in a practical CV-QKD system~\cite{CVQKD_securityproof}.  
 
\section{Comparison of communication protocols.}\label{Comp_Comm}

In this section, we provide a detailed comparison of various communication protocols using classical, quantum and post-quantum cryptography based on their security properties, key management requirements, complexity, and scalability. The discussion aims to offer a comprehensive understanding of the trade-offs involved in selecting an appropriate protocol for different security environments. All of these protocols are summarized in \cref{tab:security_comparison}.

\subsection{Pre-shared keys + \glsxtrshort{OTP} + \glsxtrshort{ITS}-\glsxtrshort{MAC}}
This method is one of the most secure but least scalable cryptographic techniques. It provides perfect confidentiality due to the \glsxtrshort{OTP} and statistical authenticity through \glsxtrshort{ITS} \glsxtrshort{MAC}. However, it lacks \glsxtrshort{PCS}, meaning that if a key is exposed, all communications encrypted with that key are compromised. A significant drawback is the need to securely distribute and store extremely large keys, as each message requires a key of equal length. Additionally, due to the requirement for in-person key exchanges, this scheme is highly impractical for large-scale networks or frequent communications. The combination of high security and logistical difficulty makes it suitable for only highly sensitive, low-frequency communications.

\subsection{Pre-shared keys + Symmetric encryption}
This approach balances security and practicality better than OTP-based methods. While it provides computational confidentiality and authenticity, it does not offer \glsxtrshort{PCS}. Symmetric encryption reduces the key size requirement compared to \glsxtrshort{OTP}, making it easier to manage. However, the challenge of securely distributing and storing keys remains. This method requires pre-shared keys, meaning a trusted mechanism must exist for exchanging them beforehand. Key management complexity is moderate to high, depending on the size of the network. Scalability is still poor since secure key exchange must occur in advance and in a quadratic fashion for a fully connected network. This protocol is occasionally used in secure messaging applications where a fixed group of users communicates regularly.

\subsection{Pre-shared authentication keys + \glsxtrshort{ITS}-\glsxtrshort{MAC} + \glsxtrshort{QKD} + OTP}
This method provides \glsxtrshort{Q-ITS} confidentiality through \glsxtrshort{OTP} encryption and statistical authenticity via \glsxtrshort{ITS} \glsxtrshort{MAC}. Additionally, it achieves \glsxtrshort{PCS}, meaning that even if an encryption key is compromised, a new one can be generated, reestablishing confidentiality for future messages, as long as the MAC-key did not become corrupted. However, deploying \glsxtrshort{QKD} requires specialized quantum infrastructure, which adds significant complexity and cost. Moreover, scalability is severely limited due to the requirement for both quantum channels and a quadratic number of in-person key exchanges. This makes the approach feasible only in very specific environments, such as in some settings involving government and military communications.

\subsection{QKD + \glsxtrshort{OTP} + Signatures}
This method offers \glsxtrshort{Q-ITS} confidentiality using \glsxtrshort{OTP} encryption and computational authenticity through digital signatures. It also ensures \glsxtrshort{PCS}, meaning security remains intact even if past encryption keys are exposed. \glsxtrshort{QKD} allows for the on-demand generation of \glsxtrshort{OTP} keys, eliminating the need for long-term key storage. The inclusion of digital signatures ensures authenticity. However, the overall complexity is high due to the reliance on quantum infrastructure. Scalability is constrained by the requirement for quantum channels, which are difficult to deploy over long distances. While star-network architectures can alleviate some of the challenges, they introduce a need for trust in the network operator.

\subsection{QKD + Symmetric encryption + Signatures}
This scheme uses \glsxtrshort{QKD} to generate encryption keys for symmetric encryption while relying on digital signatures for authentication. Unlike \glsxtrshort{OTP}-based approaches, symmetric encryption reduces the key size requirement, making storage and management easier. However, the confidentiality guarantee now depends on both computational assumptions \textit{and} \glsxtrshort{Q-ITS}. \glsxtrshort{PCS} is ensured, meaning that future communications remain secure even if past keys are compromised. The major drawbacks are the reliance on \glsxtrshort{QKD} infrastructure, which significantly increases deployment complexity and the reliance on twos sets of security assumptions (computational and quantum information theoretical) instead of just one in the cases of both QKD and solutions that are based on classical authenticated key exchanges. Scalability remains poor due to the need for quantum channels, but some network architectures, such as star topologies, can improve feasibility.

\begin{table}[h!]
\centering
\caption{Comparison of communication protocols using different combinations of classical, quantum, and post-quantum cryptography.}
\label{tab:security_comparison}
\begin{tabular}{|>{\raggedright}p{2.25cm}|>{\raggedright}p{2.25cm}|>{\raggedright}p{2.75cm}|>{\raggedright}p{2.75cm}|>{\raggedright}p{3cm}|}
\hline
\textbf{Scheme} & \textbf{Security} & \textbf{Key Management} & \textbf{Complexity} & \textbf{Scalability}\\
\hline
    \textbf{Pre-shared keys + \glsxtrshort{OTP} + \glsxtrshort{ITS} 
  \glsxtrshort{MAC}}
    & Perfect confidentiality, statistical authenticity. No \gls{PCS}.
    & Biggest challenge is securely sharing, re-sharing (if more keys are required), and storing a key as long as the messages.
    & High complexity in key management.
    High due to the need for large keys and secure storage. 
    & Very Poor, due to need for quadratic number of in person interactions. \\
\hline
    \textbf{Pre-shared keys + Symmetric encryption} 
    & Computational confidentiality and authenticity. No \gls{PCS}. 
    & Secure key distribution is required. Key size is much smaller than in OTP, making it more practical.
    & Moderate to high in key management. Slightly easier than with \glsxtrshort{OTP} due to smaller key sizes.
    & Very Poor, due to need for quadratic number of in person interactions. \\
\hline
    \textbf{Pre-shared authentication keys + \glsxtrshort{ITS} \glsxtrshort{MAC} +\glsxtrshort{QKD}+ OTP}
    & Q-ITS confidentiality, statistical authenticity. \gls{PCS}.
    & Secure key distribution is required. \glsxtrshort{QKD} allows on-demand generation of OTP-keys.
    & Moderate to high in key management, but overall high due to need for \glsxtrshort{QKD} infrastructure.
    & Extremely poor, due to quadratic number of in person interactions combined with the need for quantum channels. \\
\hline
    \textbf{QKD +\glsxtrshort{OTP}+ Signatures} 
    & Q-ITS confidentiality,  computational authenticity. \gls{PCS}.
    &\glsxtrshort{QKD} allows on-demand creation of OTP-key, signature verification key needs to be securely shared.
    & High complexity due to need for \glsxtrshort{QKD} infrastructure. Low complexity for key management.
    & Poor, due to need for quantum channels; star-networks can take some of the edge of, but inherently require trust into operator. \\
\hline
    \textbf{QKD + Symmetric encryption + Signatures}
    & Computational confidentiality with additional need of quantum-assumptions, computational authenticity. \gls{PCS}.
    &\glsxtrshort{QKD} allows on-demand creation of encryption-key, signature verification key needs to be securely shared.
    & High complexity due to need for \glsxtrshort{QKD} infrastructure. Low complexity for key management.
    & Poor, due to need for quantum channels; star-networks can take some of the edge of, but inherently require trust into operator. \\
\hline
    \textbf{Authenticated key exchange + Symmetric encryption} 
    & Computational confidentiality and authenticity. \gls{PCS}.
    & \glsxtrshort{AKE} allows on-demand creation of encryption-key, authenticity-key needs to be securely shared and stored.
    & Low complexity for key management.
    & Excellent, if \glsxtrfull{PKI} is available, otherwise still better than alternatives. \\
\hline
\end{tabular}

\end{table}

\subsection{Authenticated key exchange + Symmetric encryption}
This approach offers computational confidentiality and authenticity in a highly scalable and practical way. Most modern protocols offer \glsxtrshort{PCS} as a matter of course and generate dynamic encryption keys, eliminating the need for long-term key storage. Key management is relatively straightforward, making it a suitable choice for large networks. Scalability is excellent, particularly when combined with a \glsxtrshort{PKI}, which enables efficient and automated distribution of authentication-keys. Most modern secure communication systems, such as \gls{TLS} and \glspl{VPN}, rely on this kind of protocol for their security.

\section{Current use cases}\label{use-cases}
	This section briefly introduces the use cases being analyzed in the subsequent sections. Analyzing all the available use cases in the literature, in commercial QKD company documents~\cite{idquantiqueTechnology,TosibaCasesQuantum,thinkquantum}, in the OpenQKD project~\cite{openqkdHome,qtOPENQKDOpen} and in other sources is not possible due to the limited details provided on some use cases. Use cases with sufficient technical information are analyzed here. The order is based on the year of occurrence. We summarize the use cases in \cref{tab:use-case-summary}.

\subsection{Authenticity of election results (2007)}\label{usecase:elections}

The State of Geneva, Switzerland, used QKD during its 2007 election process to secure transmission of the election count totals from the counting center to the location where the votes were stored~\cite{id_quantique_idq_2017,id_quantique_elections_use_case_government,houston_iii_secure_2007}.
The QKD keys were used in conjunction with \glsxtrshort{AES}-256 for confidentiality and \glsxtrshort{HMAC}-\glsxtrshort{SHA}-256 for authenticity~\cite{id_quantique_cerberis_spec_2007,id_quantique_vectis_spec_2005}.
ID Quantique provided the QKD devices. The locations were directly connected using 4 km long fibers~\cite{nicolas_gisin_notitle_2007}.

\subsection{Backup for disaster recovery (2017)}\label{usecase:disasterBackup} A private asset and wealth management company in Switzerland needed to secure its communication network between its headquarters and a \glsxtrfull{DRC}~\cite{disaster-recovery-bank, cerberis-specs-2012}. To protect sensitive data long-term, they used Thales' network encryptors and Cerberis QKD devices to combine layer 2 Ethernet encryption using AES-256 with keys from QKD. The success of this implementation led to the expansion of the encryption platform to other areas of the company for \glsxtrshort{MAN} and \glsxtrshort{WAN} applications.

\subsection{Financial data (2019)}\label{usecase:toshibaFinance} In 2019, Toshiba and Quantum Xchange reported on a collaboration that augmented the encrypted connection between Wall Street's financial markets and a data center in New Jersey, using the QXC Phio QKD network~\cite{toshiba_financial,quantum_xchange_toshiba}.
The connection between Wall Street and the data center is usually used to transmit sensitive financial data like trading algorithms and customer settlement accounts.
As a demonstration, the QKD-augmented connection was used to transmit an uncompressed live video stream.
The fiber connection multiplexes the QKD channels, including the quantum one, and the commercial data over a single dark fiber. The multiplexing scheme was based on \glsxtrfull{CWDM} with the quantum channel in the O-band and the rest of the channels in the C-band. Deploying new optical fibers is one of the most expensive operations for a network provider. Therefore, the capability of Toshiba’s system of avoiding a dedicated dark fiber for the quantum channel only is a significant improvement for the solution proposed in this use case.

\subsection{Distributed information sharing and backups}

We encountered several use cases in which actors wanted to store sensitive information, which we describe below.
All use cases followed the principle of protecting that information against server breaks via a cryptographic protocol known as `Secret Sharing', i.e., by splitting them into $N$ many components (`shares') that cannot be used to reconstruct the original information without obtaining at least $n$ many shares out of these $N$.
The rationale is that by distributing the separate shares to spatially separated secure locations, it is ensured that an attacker cannot access the information without breaking into at least $n$ many (secured) servers.

\subsubsection{Facial recognition (2019)}\label{usecase:facialRecognition} The \glsxtrfull{NICT} of Japan, \glsxtrfull{NEC}, and the National Olympic Committee of Japan have partnered to use facial recognition to access a server room~\cite{nict_facial_recognition}. QKD is used to secure the necessary data for facial recognition. A facial recognition server decides if a person gets access to a server room that stores medical and athlete data records. The video recordings of athletes are used for analysis to improve their performance. A camera is connected to the central server and over this connection facial recognition data is sent which is secured using QKD. The central server is also connected to three servers using a QKD-secured connection. The biometric data of people who should be given access to the server room is stored on the facial recognition server and a backup is stored on three servers using secret sharing.

\subsubsection{Key storage and key backup (2020)}\label{usecase:vault} Mt. Pelerin, a Swiss company specializing in cryptocurrency, partnered with ID Quantique to involve QKD in their asset management to secure digital assets~\cite{key_storage,id_quantique_id_2020}\cite[Use Case 03]{ofr}. These assets include blockchain private keys with which transactions can be signed and were split into five shares using \glsxtrfull{SSS}~\cite{shamirs-secret-sharing}. To recover the assets, access is necessary to three of five storage nodes~\cite{idq_use_cases_presentation}. The procedure for backing up assets to storage nodes involves OTP encryption of each share, for which the application uses keys distributed by ID Quantique’s QKD devices.

\subsubsection{Medical image sharing 
 (2020)} \label{usecase:medicalGraz} The Diagnostic and Research Center for Molecular BioMedicine of Medical University Graz, Austria, exchanged images with the pathological institute of the LKH Graz West II, so that the images could be analyzed at both sites~\cite{secret_sharing_graz_paper, secret_sharing_graz_press}.
The images were split in three shares using fragmentiX secret sharing which is based on \gls{SSS}~\cite{fragmentix-ss}. Two shares were encrypted using AES with QKD keys and transferred to two different datacenters in the same city. The third share was transferred using TLS to a storage at the Medical University Graz. In case one storage location encounters data loss, all data can be restored using the other two storage locations. 

\subsubsection{Medical record backup (2020)}\label{usecase:medicalBackup} NICT, NEC and ZenmuTech partnered to backup medical records~\cite{nec_medical,sasaki_qkd_2017}. Dummy medical records were sent from the medical institution to a server using QKD. This server split the medical records into three shares using the \glsxtrfull{AONT}~\cite{aont}. The three shares were then transmitted to three different data centers in different cities while being secured by QKD.

\subsection{Connecting data centers (2020)}\label{usecase:geneveDatacenter} \glsxtrfull{SIG}, a Swiss public utility company managing a fiber optical network, partnered with ID Quantique (IDQ) to secure a connection between two data centers using QKD and AES~\cite[Sec. 3.4]{ofr}. SIG runs an encrypted connection between their two main data centers to secure sensitive data processed in their cloud applications~\cite{datacenter}.
They used QKD keys to augment the standard encryption key. Functionality in the case of unavailable QKD keys is maintained by falling back to the non-augmented encryption keys.

\subsection{Genome data (2020)}\label{usecase:toshibaGenome} By 2020, Toshiba Corporation and Tohoku Medical Megabank Organization (ToMMo) finished a five-year trial in which they used Toshiba’s QKD system to encrypt sensitive large-scale genome sequence data~\cite{toshiba_genome,toshiba_genome2,toshiba_global_genome_2020}.
Over two years, genome data produced with the Japonica Array tool was encrypted and transmitted from the Toshiba Life Science Analysis Center to the Tohoku Medical Megabank Organization over a distance of 7 km. Toshiba reported to have achieved stable communication with speeds exceeding 10 Mbps.

Toshiba furthermore reported that the sequencing of 24 genome data sets took over 117 hours to generate. During the generation of this data, the data was transmitted after being encrypted with OTP using keys from QKD. The transmission of this data finished in less than 4 minutes after the sequencing finished.

In 2020, the QKD network was extended with a connection between Tokohu University Hospital and ToMMo, which are a few hundred meters away from each other. The QKD keys were used to encrypt video conferences and exome sequence data using OTP. The exome sequence data was encrypted and transmitted while the sequencing was ongoing. One exome sequence produces approximately 344GB of data.

\subsection{Metrics of an overbraider machine (2020)}\label{usecase:toshibaOverbraider} Toshiba, BT, the Centre for Modelling and Simulation (CFMS), and the National Composites Centre (NCC) recently partnered to send production data between NCC and CFMS, encrypting this data using QKD-generated keys~\cite{cfms1, cfms2, overbraider-pdf}. The data, which was sent over a 7km long fiber, included quality and performance metrics of an overbraider machine.

\subsection{Self-driving cars (2021)}\label{usecase:cars}
In collaboration with QRate, researchers from Innopolis University, Russia, integrated QKD hardware into a self-driving car in 2021~\cite{self_driving_car}, to facilitate the exchange of key material with QKD-enabled gas/charging stations over an optical fiber. The gained key material was to be used to launch an encrypted OpenVPN connection (4G LTE) with the vendor's data center, in order to facilitate remote software updates and the transmission of telemetry data.

\subsection{Grid network (2022)}\label{usecase:grid} IDQ additionally partnered with SIG to connect two of SIG's power stations to the OpenQKD testbed, as a demonstrator for an envisioned Smart Grid network connecting all of SIG's power stations in Geneva with each other and the operations center by a peer-to-peer (p2p) architecture~\cite[Sec. 3.1]{ofr}. According to the OpenQKD report~\cite[Sec. 3.1]{ofr}, the use case was motivated by the goal to `secure data transmission and detect intrusions such as hackers taking control of the electricity distribution network'.

\subsection{Authentication of smart grid communications (2022)}\label{usecase:scada}

In this use case, QKD was deployed over a distance of 3.4 kilometers between a power distribution center and an electrical substation in Tennessee, United States~\cite{scada_authentication_2022}. The keys from the Qubitekk QKD system were used to authenticate \glsxtrfull{SCADA} traffic which was send using MQTT (Message Queuing Telemetry Transport). The \glsxtrshort{SCADA} traffic contains  non-confidential control data and measurement data such as voltage, current, frequency and phase. The traffic was authenticated using the \glsxtrfull{GMAC} with AES.

\subsection{Genome distance sharing (2022)}\label{usecase:qugenome}

In the QuGenome project, the \glsxtrfull{UPM}, the \glsxtrfull{CSIC} and the \glsxtrfull{Ciemat} had one or more genome sequences of which they would like to know how they are related to the genome sequences at the other institutions~\cite{qugenome_video}. Using quantum oblivious transfer, secure multiparty computation and a distance-based method, the evolutionary distances between every pair of sequences were calculated without revealing the genome sequences at one institution to the other institutions~\cite{qugenome_paper}. The evolutionary distances were encrypted using OTP with QKD keys and shared with the other institutions to calculate the phylogenetic tree of the genome sequences. This tree can be used to visualize how family members or different variants of a virus are related.

\begin{center}
\begin{landscape}
{\small
\begin{longtable}{|p{2cm}|p{2.5cm}|p{5cm}|p{5cm}|p{2.5cm}|p{3.5cm}|p{2.5cm}|}
\caption{A summary of use cases
}
\label{tab:use-case-summary}
\\ 
\hline
    \textbf{Use Case}
    & \textbf{Target Sector, Country}
    & \textbf{Description}
    & \textbf{QKD System and network}
    & \textbf{Intended impact}
    & \textbf{Security Goals}
    & \textbf{QKD service} \\
\hline
\textbf{\ref{usecase:elections}: Authenticity of election results}
    & Government, CH
    & Used QKD to guarantee authenticity of election results during transmission between the counting center and storage location.
    & ID Quantique Cerberis QKD system directly connected using a 4 km long fiber.
    & Ensured authenticity of election results.
    & Transmission reliability, successful integration of QKD with existing cryptographic methods.
    & Equipment. \\ \hline

\textbf{\ref{usecase:disasterBackup}: Backup for disaster recovery}
    & Banking, CH
    & Used QKD with AES to secure connections between a bank's headquarters and a disaster recovery center.
    & QKD via ID Quantique's Cerberis QKD server.
        \break Direct connection, approx. 100 km apart.
    & Enhanced security for disaster recovery operations.
    & Forward secrecy, high performance, secure communication.
    &  Equipment. \\ \hline

\textbf{\ref{usecase:toshibaFinance}: Financial data}
    & Financial, US
    & Used QKD to secure a video stream transmission over a 32km dark fiber between financial offices.
    & Toshiba’s QKD system. 
        \break Data transmission over 32 km dark fiber between New York and New Jersey.
    & Enhanced security for financial data transmission.
    & Increased network capacity, secure long-distance communication.
    & QKD keys as a service. \\ \hline

\textbf{\ref{usecase:facialRecognition}: Facial recognition}
    & Biometrics, JP
    & Used QKD to secure the transfer of data for facial recognition in a server room.
    & NEC's QKD devices in Tokyo QKD Network.
        \break One trusted node between the camera server and the facial recognition server.
    & Secure storage and transfer of biometric data.
    & Reliability of QKD-secured video storage, successful facial recognition data transfer.
    & Data transport. \\ \hline

\textbf{\ref{usecase:vault}: Key storage and key backup}
    & Cryptocurrency, CH
    & Used QKD with OTP to transfer key shares in a cryptocurrency exchange's key storage system.
    & QKD with ID Quantique's \glsxtrshort{QRNG}.
        \break Direct connection.
    & Secure storage and backup of keys
    & Robustness of key recovery, resilience against attacks
    & QKD keys as a service~\cite{idq_use_cases_presentation}. \\ \hline

\textbf{\ref{usecase:medicalGraz}: Medical image sharing}
    & Medical research, AT
    & Exchanged medical images using QKD and secret sharing between different datacenters.
    & QKD devices from ID Quantique, Toshiba, and ADVA.
        \break Direct connection with intermediate nodes at Citycom Graz data centers.
    & Secure storage and sharing of medical data.
    & Data recovery capabilities, redundancy of data storage.
    & Equipment. \\ \hline

\textbf{\ref{usecase:medicalBackup}: Medical record backup}
    & Healthcare, JP
    & Backed up medical records using QKD to split and transmit data to different datacenters.
    & The Tokyo QKD Network was used, which connected three locations using equipment from NEC, Toshiba, NTT-NICT and Gakushuin University.
    & Secure backup of electronic medical records.
    & Data redundancy, secure multi-location storage.
    & Data transport. \\ \hline

\textbf{\ref{usecase:geneveDatacenter}: Connecting data centers}
    & Public utility, CH
    & Secured connection between two data centers using QKD and AES-256 on layer 1 to protect cloud applications.
    & ID Quantique's Cerberis3 system.
        \break Direct connection using ADVA FSP 3000 for optical transport.
    & Increased data security for cloud applications.
    & Secure data transmission, fallback mechanisms.
    & Equipment. \\ \hline

\textbf{\ref{usecase:toshibaGenome}: Genome data}
    & Medical research, JP
    & Secured transfer of genome data and video conferences using QKD and OTP encryption.
    & Toshiba’s QKD system.
        \break Data transmission over 2 connections between 3 medical institutions, of which one connection 7 km long.
    & Secure high-speed transmission of genome data and video conferences.
    & Stable communication, real-time transmission capabilities.
    & QKD-secured service. \\ \hline

\textbf{\ref{usecase:toshibaOverbraider}: Metrics of an overbraider machine}
    & Industrial manufacturing, UK
    & Sent quality and performance metrics of an overbraider machine between two facilities using QKD.
    & Toshiba’s QKD system.
        \break Data transmission over 7 km connection.
    & Secure transmission of industrial data.
    & Secure and reliable data exchange, optimized industrial process monitoring.
    & Equipment. \\ \hline

\textbf{\ref{usecase:cars}: Self-driving cars}
    & Automotive, RU
    & Integrated QKD hardware into a self-driving car to securely transfer data and software updates over 4G LTE.
    & QRate QKD hardware.
        \break Quantum key distribution during refueling/charging via optical channel.
        \break \glsxtrshort{VPN} over 4G LTE.
    & Secure transfer of software updates and telemetry data.
    & Cache and retrieve QKD keys, secure data transfer over mobile network.
    & Equipment. \\ \hline

\textbf{\ref{usecase:grid}: Grid network}
    & Energy, CH
    & Secured data transmission between power stations using QKD to prevent intrusions.
    & QKD, Peer-to-peer architecture.
    & Enhanced security for Smart Grid communications.
    & Latency impact, link stability, service continuity during QKD-related issues.
    & Equipment. \\ \hline

\textbf{\ref{usecase:scada}: Authentication of smart grid communications}
    & Energy, US
    & Authentication of SCADA traffic (i.e. measurement and control data) between a power distribution center and an electrical substation.
    & Qubitekk QKD System. Direct connection, 3.4 km.
    & Ensured authenticity of SCADA traffic.
    & Information-theoretic authentication in smart grid communications.
    & Equipment.
    \\ \hline

\textbf{\ref{usecase:qugenome}: Genome distance sharing}
    & Medical research, ES and PT
    & Used QKD with OTP to secure information sharing on how genome sequences (for example from family members) are related.
    & ID Quantique and Huawei QKD systems were used to secure 2 connections of 7 km and 24 km between 3 institutions.
    & Secure sharing of genome distances.
    & Securely calculating and sharing of genome distances without revealing the genome sequences.
    & Equipment.
    \\ \hline

\end{longtable}

}

\end{landscape}
\end{center}

\section{Method of analysis}\label{analysis-method}
	While the use cases listed in~\cref{use-cases} differ in terms of target sector and concrete technological means, they share similarities in their technological approaches and/or their security goals. To structure our discussion in~\cref{analysis}, we will analyze the use cases using the following questions:

\begin{itemize}
    \item How is the key used?
    \item What does the network topology look like?
    \item What security guarantees does the system provide?
\end{itemize}

\subsection{How is the key used?} 

While all use cases involve the use of keys provided by QKD, these keys are used in different protocols. The respective protocol determines
\begin{itemize}
	\item which data is encrypted. For example, MACsec encrypts the source and destination IP address, while TLS does not. 
	\item and possibly, the symmetric cipher used to encrypt the data. 
\end{itemize}
Different symmetric ciphers have different security guarantees. While AES provides computational security, OTP provides secrecy independent of the computational resources of the attacker, assuming the secrecy of the key does not depend on this as well and assuming the key is only used once. This has implications for the security guarantees that are provided by the system.
To analyze the use cases in~\cref{analysis}, it is therefore important to keep the following questions in mind:

\begin{itemize}
    \item What is the key used for? What kind of data will the key protect?
    \item By which protocol is the key used? Does it involve any additional cryptography? If yes, what are the implications if that additional cryptography breaks?
    \item How long does the data need to be protected?
\end{itemize}

\subsection{What does the network topology look like?} 

Understanding network topology is crucial because it directly impacts the security, scalability, and feasibility of QKD deployment. The choice of QKD protocols, devices, and the configuration of trusted nodes influences overall security guarantees and determines the practicality of implementing QKD in real-world scenarios. By examining these aspects, we can assess how well the system aligns with the intended use case and its security requirements. Therefore, we look for the following points:
\begin{itemize}
    \item What does the network topology look like?
    \item What QKD protocols and devices are used? 
\end{itemize}

\subsection{What security guarantees does the system provide?} The two primary properties that we discuss in this work are  confidentiality and authenticity.
Additionally we require correctness (the property that the scheme `works' in the absence of an attacker) and 
availability, the property that a scheme can be used when needed. For the most part, the latter two are however less of an issue and none of the use cases that we analyzed seemed to encounter major issues with them.

\section{Use case assessment and recommendations}\label{analysis}
	To address the specific questions raised in \cref{analysis-method} for each use case, we now examine to what end the established QKD keys are used, which protocols and devices are involved, and the obtained security guarantees.
 To enable a comparison with traditional cryptographic methods and \gls{PQC}, we additionally discuss if and how traditional cryptography and \gls{PQC} could enable each use case, and if this would lead to different security guarantees. We also critically comment on the discussed use cases and offer recommendations for improvement. We summarize the analysis in \cref{tab:analysis-use-cases}.
We also provide an overview over alternatives that don’t rely on QKD to achieve the same security goals. We remark that we do so largely independently of the needed assumptions, in particular we assume that quantum assumptions and computational assumptions are both suitable to meet the security requirements.

\subsection*{Comment on backups in general}\label{SA:commentBackups}

We encountered several use cases that use QKD to protect the transmission of data with the goal to backup this data.  QKD is primarily a replacement for key exchange mechanisms aiming for transport protection.

However, transported data should ideally also be protected at rest, i.e., while being stored as backups. Generally, this is not an ideal use case for dynamically created (or `non-static') keys since such keys have to be stored along with the backup for eventual data recovery.
For backup use cases, it may be desirable to use a pre-existing key to encrypt the to-be-transmitted data already before transmission since the storage facility then does not have to store the key material, making it harder for attackers to recover the encrypted data even if they gain digital or physical access to the storage facility.
If the encryption scheme used for this long-term encryption is asymmetric, the generator of the data does not even need to maintain knowledge about the used secret key -- it could be stored in a separate secure location, independent of the mass-data backup.

One way to then further strengthen such storage-encryption is to secret-share the key in a way that requires multiple parties to work together to decrypt the back-ups. Depending on the use-case this could for example involved all members of a company's board of directors receiving a share.

\subsection{Authenticity of election results (\ref{usecase:elections})} \label{SA:elections}
\subsubsection*{Aim}
\begin{itemize}
    \item Secure the link between the central ballot-counting station in downtown Geneva and government data centers in the suburbs over fiber-optic channels.
    \item The QKD key ensures the authenticity of election results during transmission.
\end{itemize}

\subsubsection*{Used technology}
\begin{itemize}
    \item \textit{Protocol and devices:} ID Quantique Cerberis QKD using SARG04 and ID Quantique Vectis link encryptor using AES-256.
    \item \textit{Connection:} Direct end-to-end connection using 4 km long fibers.
    \item \textit{Additional Crypto:} AES-256 for confidentiality of election results, HMAC-SHA-256 for authenticity of election results.
\end{itemize}

\subsubsection*{Analysis and recommendation}

In this use case, the QKD key was used to protect count totals of a public election. This data can be made public as soon as the voting stations close. If the data is transferred after closing the voting station, it does not have to be confidential anymore. The data only has to be authenticated to make sure the count totals are not modified. 
Therefore, it is not clear that this use case requires any form of confidentiality whatsoever; it seems that it only requires authenticity.
Any QKD protocol requires an authenticated channel, which would suffice to solve the use case in a more natural way (by directly authenticating the transmitted local results once the election concluded).
Therefore, we assess that QKD does not add value to this use case.

One could envision a variant of this use case in which votes are transmitted as they are being cast, instead of being transmitted only as an aggregate that needs no confidentiality.
This would change the setting to a setting that requires confidentiality, which could be provided by QKD or AKEs. This variant, however, would inherently introduce additional attack vectors -- e.g., parties with access to the receiving system could collaborate with a party that collects information about when a given voter voted, thereby breaking the secrecy of the election.

\subsection{Backup for disaster recovery (\ref{usecase:disasterBackup})}
\label{SA:disasterBackup}
\subsubsection*{Aim}
\begin{itemize}
    \item Protect highly sensitive business and customer data, including financial records, credit card information, personal details, and any other critical data relevant to the company's operations and customer trust.
    \item The established QKD key is used to maintain security in case the pre-shared key is compromised. The QKD key is combined with the AES-256 pre-shared encryption key used by the 10 Gigabit Ethernet encryptors to ensure forward secrecy and protect against eavesdropping and future decryption attempts. The combined key is used  for data transmitted between the company's headquarters and the \glsxtrshort{DRC}.
\end{itemize}

\subsubsection*{Used technology}
\begin{itemize}
    \item \textit{Protocol and devices:} ID Quantique's Cerberis QKD system using SARG04 and Thales' encryptor using AES-256.
    \item \textit{Connection:} The setup includes a direct end-to-end connection between the headquarters and the \glsxtrshort{DRC}, approximately 100 kilometers apart. The connection uses four 10 Gigabit Ethernet encryptors that facilitate the encrypted data transmission.
    \item \textit{Additional Crypto:} AES-256 to protect business and customer data.
\end{itemize}

\subsubsection*{Analysis and recommendation}\label{SA:recBackups}
QKD is used here to provide transport security, which is it’s primary strength. However, the backup data is only secured in transit and not secured at rest. The recommendation to secure the data at rest from the \nameref{SA:commentBackups} applies.

\subsection{Financial data (\ref{usecase:toshibaFinance})} \label{SA:toshibaFinance}

\subsubsection*{Aim}
Secure the transmission of sensitive and high-value data between a datacenter on Wall Street and a datacenter in New Jersey. This data includes trading algorithms, customer settlement accounts, real-time trading and transactional data, core banking applications and video conferencing data.

\subsubsection*{Used technology}
\begin{itemize}
    \item \textit{Protocol and devices:} Toshiba’s QKD system using BB84/T12 and Senetas encryptor using AES.
    \item \textit{Connection:} Toshiba’s Multiplexed Single Fiber QKD system operating in the O-Band is used. This system allows combined commercial data and QKD traffic over a single fiber. The data transmission occurs over a 32 km dark fiber between New York and New Jersey.
    \item \textit{Additional Crypto:} AES-256 to protect financial data.
\end{itemize}

\subsubsection*{Analysis and recommendation}

QKD is used here to provide transport security, which is it’s primary strength. We only note that the institutions in question are physically close enough and pre-determined enough, that exchanging AES-keys physically, as discussed in \cref{background:OTPwithPSK}, would be a viable and even more secure, but more labor-intensive alternative here.

\subsection{Distributed information sharing and backups} \label{SA:backups}

\subsubsection{Facial recognition (\ref{usecase:facialRecognition})} \label{SA:facialRecognition}

\subsubsection*{Aim}
Secure the transmission of biometric authentication data, specifically feature data in a face recognition system, between the central server and the face recognition server. Additionally, secret sharing is used to secure the storage of reference data for authentication across distributed servers. 

\subsubsection*{Used technology}
\begin{itemize}
    \item \textit{Protocol and devices:} NEC's QKD devices, using BB84 and \glsxtrshort{DPS-QKD}, integrated into the Tokyo QKD Network.
    \item \textit{Connection:} There is one trusted node between the camera server in the NOC (Network Operation Center) in Tokyo and the facial recognition server at NICT headquarters.
    \item \textit{Additional Crypto:} \glsxtrshort{SSS} to split biometric data and an unknown symmetric cipher to encrypt the split biometric data.
\end{itemize}

\subsubsection*{Analysis and recommendation}

As biometric authentication cannot be changed like a password, this data has to be protected for a long time. QKD is indeed a possible way to provide some transport security, though AKEs would be able to do the same and the \nameref{SA:commentBackups} also applies to this use case.

\subsubsection{Key storage and key backup (\ref{usecase:vault})} \label{SA:vault}
\subsubsection*{Aim}
\begin{itemize}
    \item Protect the split private keys that correspond to digital assets such as cryptocurrencies. These private keys are critical as they are the proof of ownership of the digital assets.
    \item The established QKD key is used for encrypting the private keys that have been split using \glsxtrfull{SSS}. Specifically, each component is encrypted with a OTP, and the keys necessary for this OTP encryption are distributed via QKD.
\end{itemize}

\subsubsection*{Used technology}
\begin{itemize}
    \item \textit{Protocol and Devices:} ID Quantique's \glspl{QRNG} are used to generate the blockchain private keys on a \glsxtrfull{HSM}. The private keys are then secured in transit using \gls{QKD} keys from ID Quantique's \gls{DV-QKD} devices.
    \item \textit{Connection:} Management node connected with five storage nodes using QKD.
    \item \textit{Additional Crypto:} To sign transaction on a blockchain, a public-private key pair is generated. OTP is used to encrypt the secret shares which were split using \glsxtrshort{SSS}.
\end{itemize}

\subsubsection*{Analysis and recommendation}

The private keys protected by the system need to be protected for the entire duration that the digital assets are in custody, which could potentially be indefinite or until the assets are moved or redeemed.
However, assuming these private keys are used to control cryptocurrencies such as Bitcoin or Ethereum, the private keys can be calculated from the public key by a \glsxtrshort{CRQC} and the usage of QKD in this use case will not protect the private key from being recovered by a \gls{CRQC}.

The use case aims to protect asymmetric secrets, i.e., secrets that are used by asymmetric cryptosystems. The use case thus assumes that the respective cryptosystems are secure -- otherwise, protection of these secrets would not be worthwhile. Considering the significant overhead of QKD in comparison with classical cryptosystems, this calls the use case into question in general.

\subsubsection{Medical image sharing (\ref{usecase:medicalGraz})} \label{SA:medicalGraz}

\subsubsection*{Aim}
\begin{itemize}
    \item Protect sensitive medical data, including digital histological slides (up to 10 GB per image), clinical and genetic data, and other medical records and images exchanged between Medical University Graz and Hospital Graz II. 
    \item The established QKD key is used for encrypting split data fragments during transmission between the hospitals (Medical University Graz and Hospital Graz II) and external S3 storage locations. The data was split using fragmentiX secret sharing.
\end{itemize}

\subsubsection*{Used technology}
\begin{itemize}
    \item \textit{Protocol and Devices:} ID Quantique DV-QKD system and Toshiba BB84/T12 QKD system. The data was encrypted using AES with an encryptor provided by ADVA.
    \item \textit{Connection:} The two hospitals were directly connected to the two datacenters operated by Citycom Graz. The datacenters function as an intermediate node for the connection between the hospitals.

    \item \textit{Additional Crypto:} Additional cryptographic measures include fragmentiX Secret Sharing (which is based on \gls{SSS}) and traditional TLS protection for one of the data fragments.
    
\end{itemize}

\subsubsection*{Analysis and recommendation}

This use case uses QKD to provide transport security, which is its primary strength. Given the nature of medical records, protection may be required for many years or even decades. Even if the TLS connection used to transport shares of the medical images is broken by an attacker, the attacker would still need access to a second share to restore the original image. Besides the caveats regarding the use of QKD in general, we note that the institutions in question are physically close enough (and pre-determined enough) that exchanging AES keys physically, as discussed in \cref{background:OTPwithPSK}, would be an alternative that is viable and does not have to rely on quantum assumptions.

\subsubsection{Medical record backup (\ref{usecase:medicalBackup})} \label{SA:medicalBackup}

\subsubsection*{Aim}
Secure the transmission of medical records that were split using secret sharing. Medical records contain detailed patient information, diagnostic results, treatment plans, and other confidential health-related data. 

\subsubsection*{Used technology}
\begin{itemize}
    \item \textit{Protocol and Devices:} The system utilizes QKD equipment from the Tokyo QKD Network, which uses equipment from NEC (BB84), Toshiba (BB84/T12), NTT-NICT (\gls{DPS-QKD}) and Gakushuin University (\gls{CV-QKD}).
    \item \textit{Connection:} A medical institution is connected using QKD to a secret sharing server. This secret sharing server is connected using three QKD connections to three servers.
    \item \textit{Additional Crypto:} \gls{AONT}, which uses AES~\cite{zenmutech_about_2024}, to split medical records and an unknown symmetric cipher to encrypt the split medical records.
\end{itemize}

\subsubsection*{Analysis and recommendation}

The confidentiality of the medical data needs to be protected for as long as the data is sensitive, which generally means many years. The data is sent to a server that secret-shares them and distributes the shares towards storage servers. As a consequence, the connections between the sender and the various servers have to be secured.
Considering that the medical data is secret shared by a trusted server and not the original producer raises significant questions about why this is done this way, as it introduces the need for an otherwise unnecessary third party. Getting rid of this component and encrypting the data directly under secret-shared keys could improve the overall security and would still allow to use QKD to protect transport-encryption as a defense-in-depth measure. At that point the analysis given in \cref{SA:medicalGraz} would apply.

\subsection{Connecting data centers (\ref{usecase:geneveDatacenter})}
\label{SA:geneveDatacenter}
\subsubsection*{Aim}

Secure the connection between SIG’s two main data centers to protect utility operations data, customer information, and other confidential business data.   

\subsubsection*{Used technology}
\begin{itemize}
    \item \textit{Protocol and Devices:} ID Quantique's Cerberis3 using the \glsxtrshort{COW} protocol and ADVA's FSP3000 encryptor using AES-256-GCM.
    \item \textit{Connection:} There is a direct end-to-end connection between the two main data centers using ADVA’s FSP 3000 product for optical transport. The setup does not mention intermediate nodes explicitly, suggesting a direct link.
    \item \textit{Additional Crypto:} AES-256-GCM to protect confidential business data. The QKD key is \glsxtrshort{XOR}ed with the standard session key to generate a super session key, which is then used by the ADVA FSP 3000 encryption equipment for Layer 1 encryption of data in transit over the optical network~\cite{idq_use_cases_presentation}
\end{itemize}

\subsubsection*{Analysis and recommendation}

QKD is used to provide transport security, its primary strength. We note, however, that this use case is highly vulnerable to downgrade-attacks -- attackers can completely eradicate the QKD component from this solution by simply interrupting the quantum channel. When the quantum channel stops working, no QKD keys will be used and only the standard session key is used to encrypt data. This behaviour prevents the unavailability of the connection.  We recommend to change this behaviour.
We also note that \glsxtrshort{XOR}ing keys is not without issues if it cannot be fully guaranteed that the keys are independent and non-maliciously generated; a dual-PRF could be more appropriate here~\cite{kem-combiners}.

We were unable to find information about the distance between the data centers. In case they are physically close, the alternative mentioned in \ref{SA:medicalGraz} (physical exchanges of key material) would also apply here.

\subsection{Genome data (\ref{usecase:toshibaGenome})}
\label{SA:toshibaGenome}
\subsubsection*{Aim}

Secure the transmission of video conferences, exome data and genome data, which can legally be considered personally identifiable information.

\subsubsection*{Used technology}
\begin{itemize}
    \item \textit{Protocol and Devices:} Toshiba’s QKD system using BB84/T12.
    \item \textit{Connection:} Data transmission occurs over 2 connections. One connection with a distance of 7 km between Toshiba Life Science Analysis Center and \glsxtrfull{ToMMo}, and one connection between \glsxtrshort{ToMMo} and Tokohu University Hospital which are a few hundred meters away from each other. \glsxtrshort{ToMMo} can function as a trusted node.
    \item \textit{Additional Crypto:} One-time pad encryption to encrypt exome and genome data.
\end{itemize}

\subsubsection*{Analysis and recommendation}

QKD is used to its primary strength, providing transport security. This transport security is used to protect genome and exome sequence data, which is sensitive personal information that should be protected for a long time. Different from most other use cases, the one-time pad is used instead of \gls{AES}. This prevents the need to rely on computational hardness assumptions. Considering the short distance between the endpoints, the comment about physical exchanges of key material (see \ref{SA:medicalGraz}) also applies here. Amongst the use cases we analyzed, we view this one and \ref{SA:qugenome} as the two more reasonable ones due to the lack of obvious and/or fundamental problems (in comparison with the other use cases) and because of the use of one-time pad encryption instead of AES encryption.

\subsection{Metrics of an overbraider machine (\ref{usecase:toshibaOverbraider})}
\label{SA:toshibaOverbraider}
\subsubsection*{Aim}
\begin{itemize}
    \item Secure the transmission of manufacturing production data between the National Composites Centre (NCC) and the Centre for Modelling \& Simulation (CFMS).
    \item Secure data from the NCC's Overbraider machine, which weaves carbon fiber to create precision hollow composite components, such as aircraft engine blades. This data is critical for the manufacturing processes and may include detailed production parameters, measurements, and assessments necessary for remote monitoring and control of manufacturing operations.
\end{itemize}

\subsubsection*{Used technology}
\begin{itemize}
    \item \textit{Protocol and Devices:} Toshiba’s QKD system using BB84/T12.
    \item \textit{Connection:} The system uses BT Openreach’s standard fiber optic infrastructure and includes QKD enabled encryption tunnels between two Edge firewalls.  The data transmission occurs over a dedicated 7 km long point-to-connection between the NCC and CFMS sites.
    \item \textit{Additional Crypto:} Unknown symmetric cipher to protect the transmission of manufacturing production data.
\end{itemize}

\subsubsection*{Analysis and recommendation}

We arrive at similar conclusions as in \cref{SA:medicalGraz}, in that we don’t see anything fundamentally wrong about the way that QKD is used here.

\subsection{Self-driving cars (\ref{usecase:cars})} \label{SA:cars}
\subsubsection*{Aim}
\begin{itemize}
    \item Secure an OpenVPN connection, securing software updates and telemetry data of an autonomous control system of a driverless car.
    \item Secure the real-time transmission of telemetry data on the status of all its subsystems to the laboratory monitoring system.
    \item Secure software updates upon the release of new versions, preventing unauthorized access or alteration of the software update.
\end{itemize}

\subsubsection*{Used technology}
\begin{itemize}
    \item \textit{Protocol and Devices:} QRate QKD system using BB84~\cite{Qrate}.
    \item \textit{Connection:} QKD between the driverless car and the data center during refueling or charging for an electric vehicle which occurs via an optical channel.
    \item \textit{Additional Crypto:} OpenVPN over 4G LTE to protect software updates and telemetry data.
\end{itemize}

\subsubsection*{Analysis and recommendation}

In the case of rental cars, we note that the owning company will be in regular physical contact with them for maintenance.
In this scenario, installing pre-shared keys appears to be an alternative that uses much simpler and cheaper technology, while at the same time being significantly more secure since it does not have to rely on trusted nodes (recall \cref{background:QKDassumptions}).
 
In the case of privately owned cars, we first note that transmitting significant amounts of real-time telemetry data to the manufacturer could pose a significant infraction of privacy, which would make transmission undesirable in any case.
Even when setting these ethical concerns aside, it is still not clear why this would make a convincing use case: it would still be viable and more practical to rely on pre-installed  key material that gets updated (replaced) during the necessary regular maintenance in a car workshop.

For the protection of software updates, we note that QKD is used purely to ensure authenticity. Like in the voting use case (\ref{SA:elections}), this begs the question from where the QKD protocol derives its authenticated channel and why that authentication mechanism couldn’t be used directly for the software updates instead.

\subsection{Grid network (\ref{usecase:grid})} \label{SA:grid}

\subsubsection*{Aim}

\begin{itemize}
    \item Secure Geneva's smart grid network (800+ power stations).
  \item Test and assess QKD technology in a real operational environment.
\item Prevent hacking and ensure secure data transmission between power stations.
\end{itemize}
    
\subsubsection*{Used technology}
\begin{itemize}
    \item \textit{Protocol and devices:} ID Quantique's DV-QKD system for QKD and Cisco's IOS-XE equipment to encrypt data in transit.
\item  \textit{Connection:} Direct connection using a dedicated 3.4 km long dark fiber for QKD and classical channels.
\item \textit{Additional Crypto:} SKIP protocol and unknown symmetric cipher to encrypt data in transit.
\end{itemize}

\subsubsection*{Analysis and recommendation}

The security goals of this use case seem underspecified -- for example, the level of protection did not become fully clear: it neither became clear what kind of data it aims to protect, nor for how long, nor against which kinds of attacks.
The stated security goals furthermore include phrasing that suggests that they are independent of the used cryptography, such as `hackers taking control of the electricity distribution network'~\cite[Sec. 3.1]{ofr} (which would go far beyond dealing with cryptography).
For network takeovers, the main attack surfaces (code execution and privilege escalation) cannot be fixed on the level of cryptographic protocols.
Our primary recommendation for this use case thus is to first create a clear threat model, to analyze which attacks follow from that threat model, and then to analyze which technologies can prevent these attacks.
In case the involved data needs long-term confidentiality (e.g. private information on energy usage),
then QKD could be an (albeit more expensive) alternative to PQC to accomplish that.

\subsection{Authentication of smart grid communications (\ref{usecase:scada})}\label{SA:scada}

\subsubsection*{Aim}

\begin{itemize}
   \item Achieve information-theoretic authentication in smart grid communications.
   \item Authenticate SCADA traffic which contains non-confidential control and measurement data.
\end{itemize}
    
\subsubsection*{Used technology}
\begin{itemize}
    \item \textit{Protocol and devices:} Qubitekk's QKD system using an entanglement-based QKD protocol. \item  \textit{Connection:} Dedicated dark fiber for QKD and classical channels.
    \item \textit{Additional Crypto:} GMAC with AES to authenticate \gls{SCADA} traffic.
\end{itemize}

\subsubsection*{Analysis and recommendation}

Although the presentation in~\cite{scada_authentication_2022} suggests that the solution achieves information-theoretic authentication, this is not the case. The solution uses GMAC with AES which is not information-theoretic secure. (GMAC with OTP would achieve statistical authenticity, but this drops to computational security when used with AES.)

This use case only requires authentication. With the same reasoning as for the other use cases that only require authenticity, voting (\ref{SA:elections}) and software updates of self-driving cars (\ref{SA:cars}), we assess QKD as not adding any value to this use case.

\subsection{Genome distance sharing (\ref{usecase:qugenome})}
\label{SA:qugenome}
\subsubsection*{Aim}
Secure the transmission of genome distance data.

\subsubsection*{Used technology}
\begin{itemize}
    \item \textit{Protocol and devices:} ID Quantique's DV-QKD system and Huawei's CV-QKD system.
    \item \textit{Connection:} Data transmission was achieved over 7 km long fibers between UPM and Ciemat and 24 km long fibers between Ciemat and CSIC. Ciemat functioned as a trusted node for the connection between CSIC and UPM.
    \item \textit{Additional Crypto:} Secure multiparty computation, which used quantum oblivious transfer, to compare genome data. One-time pad encryption to secure the transmission of genome distance data.
\end{itemize}

\subsubsection*{Analysis and recommendation}

We come to the same conclusions as for the genome distance sharing use case (\ref{SA:toshibaGenome}) with two additional remarks:
\begin{itemize}

\item As noted by the authors of the implementation, the current implementation is missing authentication for the one-time pad encryption~\cite{genome-distance-sharing-todo-encryption}.

\item Pre-shared keys could be used as an alternative. In the case that 3 institutions all have 10 SARS-CoV-2 genome sequence and would like to calculate the phylogenetic tree as given by the example in~\cite{qugenome_paper}, $ 18.6 * 10^3 $ bits of key material would be necessary to share the distances between genome sequences. Although QKD might already be available in case quantum oblivious transfer is used, instead of using QKD for the sharing of genome distances, the institutions could share SD cards with 1 \glsxtrshort{TB} of key material with each other. This would be enough key material to calculate a phylogenetic tree more than 2 million times every day for a year\footnote{A 1 \glsxtrshort{TB} SD card can provide enough keys to calculate a phylogenetic tree $(10^{12} \cdot 8)/((18.6 \cdot 10^3)/2)/365.25 = $ \num{2355140} times, every day.} and could thus be used as an alternative to QKD in this use case.
\end{itemize}

\begin{center}
\begin{landscape}
\begin{longtable}{|p{3cm}|p{9cm}|p{3cm}|p{3.5cm}|p{3cm}|}
\caption{Security analysis of QKD Use Cases}\label{tab:analysis-use-cases} \\
\hline
    \textbf{Use case}
    & \textbf{Aim}
    & \textbf{Used technology}
    & \textbf{Security Requirements}
    & \textbf{Alternatives} \\
    \hline

\textbf{\ref{SA:elections} Authenticity of election results}
    & Ensures authenticity of election results during transmission.
    \break Secures the link between central ballot-counting stations and government data centers.
    & SARG04, AES-256, HMAC-SHA-256
    & Authenticity
    & PQ Signatures
    \\ \hline

\textbf{\ref{SA:disasterBackup} Backup for disaster recovery}
    & Additional security layer for encryption between headquarters and DRC.
        \break Protects sensitive business and customer data.
    & SARG04, AES-256
    & \makecell{Confidentiality, \\ Authenticity}
    & Pre-shared keys \break \glsxtrshort{PQ} key exchange
    \\ \hline

\textbf{\ref{SA:toshibaFinance} Financial data}
    & Secures transmission of sensitive data between financial centers. Protects trading algorithms, customer accounts, etc.
    & BB84/T12, AES-256
    & \makecell{Confidentiality, \\ Authenticity}
    & Pre-shared keys \break \glsxtrshort{PQ} key exchange
    \\ \hline

\textbf{\ref{SA:facialRecognition} Facial recognition}
    & Secures transmission of biometric authentication data.
        \break Protects feature data in a face recognition system.
    & BB84 and \glsxtrshort{DPS-QKD}, Secret sharing, symmetric cipher used unknown
    & \makecell{Confidentiality, \\ Authenticity}
    & Pre-shared keys \break \glsxtrshort{PQ} key exchange
    \\ \hline

\textbf{\ref{SA:vault} Key storage and backup}
    & Encrypts digital asset components split using secret sharing.
        \break Protects private keys for digital assets such as cryptocurrencies. The to-be-protected assets are secrets for computationally secure cryptography
    & DV-QKD, OTP, Secret sharing
    & \makecell{Confidentiality, \\ Authenticity}
    & Pre-shared keys 
        \break PQ key exchange
    \\ \hline

\textbf{\ref{SA:medicalGraz} Medical image sharing}
    & Encrypts data fragments during transmission between hospitals and external storage.
        \break Protects highly sensitive medical data.
    & DV-QKD and BB84/T12, AES, Secret sharing, TLS
    & \makecell{Confidentiality, \\ Authenticity}
    & Pre-shared keys \break \glsxtrshort{PQ} key exchange
    \\ \hline

\textbf{\ref{SA:medicalBackup} Medical record backup}
    & Secures transmission of medical records.
        \break Protects extremely sensitive personal data.
    & BB84, BB84/T12, \glsxtrshort{DPS-QKD}, \glsxtrshort{CV-QKD}, Secret sharing
    & \makecell{Confidentiality, \\ Authenticity}
    & Pre-shared keys \break \glsxtrshort{PQ} key exchange
    \\ \hline

\textbf{\ref{SA:geneveDatacenter} Connecting data centers}
    & Secures symmetric key exchange between data centers.
        \break Protects utility operations and customer data.
    & \glsxtrshort{COW}, AES-256
    & \makecell{Confidentiality, \\ Authenticity}
    & Pre-shared keys \break \glsxtrshort{PQ} key exchange
    \\ \hline

\textbf{\ref{SA:toshibaGenome} Genome data}
    & Secures transmission of genome data and video conferences.
        \break Protects highly sensitive genome data and video conferences.
    & BB84/T12, OTP
    & \makecell{Confidentiality, \\ Authenticity}
    & Pre-shared keys \break \glsxtrshort{PQ} key exchange (losing everlasting confidentiality)
    \\ \hline

\textbf{\ref{SA:toshibaOverbraider} Metrics of an overbraider machine}
    & Secures transmission of manufacturing data.
        \break Protects production parameters of composite components.
    & BB84/T12, symmetric cipher used unknown
    & \makecell{Confidentiality, \\ Authenticity}
    & Pre-shared keys \break \glsxtrshort{PQ} key exchange
    \\ \hline

\textbf{\ref{SA:cars} Self-driving Cars}
    & Secures software updates and telemetry data.
        \break Quantum-protected software updates for the autonomous control system.
    & BB84, OpenVPN
    & Authenticity for Software Updates, Confidentiality and Authenticity for Telemetry.
    & Pre-shared keys \break \glsxtrshort{PQ} key exchange \break Signatures (auth only)
    \\ \hline

\textbf{\ref{SA:grid}: Grid network}
    & Secure Communication between power stations, prevent hacking
    & DV-QKD, Peer-to-peer architecture, symmetric cipher used unknown
    & Authenticity,\break Maybe Confidentiality (unclear)\break (Secure implementations; not really a crypto-problem in the first place)
    & Pre-shared keys \break \glsxtrshort{PQ} key exchange
    \\ \hline

\textbf{\ref{SA:scada}: Authentication of smart grid communications}
    & Ensure authenticity of measurement and control data
    & Entanglement-based QKD, GMAC with AES
    & Authenticity
    & Pre-shared keys \\ \hline

\textbf{\ref{SA:qugenome}: Genome distance sharing}
    & Secures transmission of genome hamming distances
    & CV-QKD, OTP, quantum oblivious transfer, secure multiparty computation
    & \makecell{Confidentiality, \\ Authenticity}
    & Pre-shared keys
    \\ \hline

\end{longtable}
\end{landscape}
\end{center}
 
\section{Conclusion}\label{conclusion}
    As quantum computing continues to develop, there is increasing attention on the risks that surround traditional cryptography in the presence of quantum attackers.
To mitigate these risks, it is necessary to consider alternative approaches such as \gls{QKD} and \gls{PQC}.
In this paper, we conducted an in-depth security evaluation of QKD-based approaches across various real-world use cases.
Our analysis highlights both the theoretical strength of QKD (not having to rely on computational assumptions), as well as its theoretical and practical limitations, such as the need to rely on physical assumptions, high implementation costs in currently available systems, poor scaling behaviour, and limitations in applicability. 

We compared QKD with other cryptographic alternatives, particularly with PQC, which offers a more direct transitioning path for existing systems and infrastructures. While PQC addresses the potential threats from quantum computers, QKD in conjunction with \glspl{OTP} might be able to offer everlasting confidentiality.
We analyzed sufficiently documented use cases and saw that in most, QKD provides very limited or no advantage over other methods for key establishment such as PQC or pre-shared keys.
The analyzed examples included using QKD for authentication, using QKD with AES to secure data between just two specific locations over a short distance, and using QKD to secure digital signing keys. The only use cases where we saw that QKD might be able to provide an advantage are use cases that use QKD with OTP (instead of AES) to secure a short-distance connection, as this might be able to secure data against an attacker with unlimited computational resources. These use cases include sharing genome data using OTP. One other potential advantage of \gls{QKD} is that it can provide \gls{PCS} without relying on computational assumptions, although none of the analyzed use cases even mention this as a security requirement.

Decision-makers must weigh the trade-offs between QKD and PQC to select the most appropriate solution for securing their systems in the post-quantum era.
 
\appendix
\section*{Appendix}

\section{Not considered use cases}

All the use cases analyzed in this paper have been sourced from various research articles, company websites, and other references. In \cref{analysis-method}, we outline the key information required for analyzing each use case. However, for some use cases, relevant data or information is unavailable for unknown reasons. To ensure completeness, we have listed these cases here. Since the necessary data is lacking, we are unable to assess their potential impact.

\subsection{Madrid QCI use cases}
Besides not finding enough information for the use cases demonstrated by the Madrid QCI~\cite{martin2023madqci}, we sometimes had other reasons for exclusion. We briefly discuss the use cases below:

\begin{itemize}
    \item \textbf{Network security and attestation} The proof of transit protocol in this use case was proven to be insecure~\cite{huitema_secdir_2021}.
    \item \textbf{Critical Infrastructure Protection} While not enough information was available to analyze this use case, it seems comparable to \ref{usecase:scada}.
    \item \textbf{QKD as a Cloud Service} Providing QKD as a cloud service is not a use case on itself, but rather a way to provide QKD.
    \item \textbf{e-Health services} Not enough information available to analyze the use case.
     \item \textbf{Quantum Cryptography for B2B and 5G} Not enough information available to analyze the use case.
     \item \textbf{Self-healed network management} Not enough information available to analyze the use case.
    \item \textbf{Quantum Cryptography with minimal amount of QKD devices allowing independent protection of users in collocated computing centers} Describes a feature of a QKD system, but not a use case.
    \item \textbf{Security independence of a network provider from QKD device manufacturers} Describes a feature, not a use case.
    \item \textbf{Open Call KaaS: Key as a Service} Providing QKD as a service is not a use case on itself, but rather a way to provide QKD.
    \item \textbf{OpenCall QGeKO} Satellite use cases are out of scope for this paper.
\end{itemize}

\subsection{Quantum CTek use cases}

The use cases listed on the website of Quantum CTek~\cite{quantuminfoTypicalCases_QuantumCTekx2013} did not include enough technical information for analysis and we therefore have been unable to verify the claims made by the original project. However, we still summarize the use cases including the unverified claims below.
\begin{enumerate}
    \item \textbf{Hefei Metro Area Quantum Communication Demonstration Network}
The Hefei Metro Area Quantum Communication Demonstration Network is the world’s first large-scale quantum network, developed and constructed by QuantumCTek. Utilizing commercial optical fiber from radio and television operators, the network covers the main urban area of Hefei city. It serves a wide range of customers, including government departments, financial institutions, military enterprises, and research institutes. The network provides quantum-secure real-time voice and text communication, file transfer, and other functions. Its topology mirrors that of the operators, featuring a three-tier structure: core, convergence, and access layers. It boasts scalability, routing capabilities, attack alarms, and a monitoring and management system that allows administrators to remotely monitor network status, observe equipment operations, and perform timely diagnostics.

\item \textbf{Financial Information Quantum Communication Verification Network of Xinhua News Agency}
This network connects the head office building of Xinhua News Agency with its Financial Information Exchange using commercial optical fiber. It provides highly confidential video and voice communication, real-time text interaction, and high-speed data transmission. The network has successfully verified a trading system based on quantum cryptography, achieving simulated equity and bond transactions.

\item \textbf{Communication Support System for a Major Event}
Designed for a significant event, this system includes a quantum telephone network and a high-speed data transmission line with quantum-safe encryption. It uses QOS-AT networking products to distribute keys without intermediate nodes. Hot backup technology ensures uninterrupted data transmission even in the event of a line fault.

\item \textbf{Quantum Secure Communication Project for Public Security}
This project establishes a standard quantum network with a cycled backbone centered around a centralized control station. Quantum gateways at each end-user node enable network expansion. As the quantum secure communication system scales, terminal nodes can be upgraded to centralized control stations, facilitating the transition to a larger quantum network.

\item \textbf{Jinan Quantum Communication Demonstration Network}
Building on the Hefei metro area's technological achievements, the Jinan network promises a secure, reliable, and integrated QKD system. At its inception, it featured the world's largest number of quantum nodes, users, business categories, and distributed keys for a metro area quantum communication network. The network supports various sectors, including government, finance, scientific research, and education, providing high-quality user experiences.

\item \textbf{Banking Regulatory Information Acquisition Demonstration}
Under the guidance of the China Banking Regulatory Commission, this network facilitates data acquisition among a supervisory unit headquarters, a local supervisory unit, and a commercial bank using a quantum secure communication network. It supports practical information submission and management, allowing banks to securely transmit data to supervisory units.

\item \textbf{Quantum Secure Communication `Beijing-Shanghai Backbone' Project}
This project aims to build a quantum secure communication backbone connecting Beijing and Shanghai via Jinan and Hefei, spanning over 2,000 kilometers. It links metro area access networks across various cities, creating a wide-area optical fiber quantum communication network. The project serves as a platform for large-scale quantum communication technology verification, research, and application demonstrations.

\item \textbf{Electronic Record Application Database Synchronization for a Large State-owned Bank}
Guided by the China Banking Regulatory Commission, a large state-owned bank uses quantum communication technology to regularly synchronize archived data from its main database to a standby database across the city. This ensures data synchronization for remote databases.

\item \textbf{Alibaba Quantum Secure Domain}
By establishing multiple quantum secure domains within the Alibaba Cloud network, Alibaba achieves inter-city data center interconnection via Quantum Portal. This setup provides customers with secure data transmission services, including key distribution, confidentiality, and networking, verified through actual business testing.

\item \textbf{ICBC Offsite Data Quantum Encryption Transmission}
Using QuantumCTek's platform, the Industrial and Commercial Bank of China (ICBC) has successfully implemented quantum communication technology for the Beijing-Shanghai offsite wide area network under the `two cities and three centers' framework. This marks the first application of quantum communication technology over a thousand kilometers in the banking industry, enabling secure online banking data transmission.

\item \textbf{Mybank Cloud Quantum Encryption Communication}
Mybank utilizes QuantumCTek’s quantum communication technology for long-distance cloud quantum encryption communication of credit business data on dedicated cloud channels between metro areas.

\item \textbf{Bank of Communications Enterprise Online Banking Use Case}
In February 2017, the Bank of Communications implemented a use case for enterprise online banking based on QuantumCTek’s platform. This marked the first application of quantum secure communication technology in real-time trading for financial enterprise online banking users.

\item \textbf{Quantum Technology in Metro Ring Network of Beijing Rural Commercial Bank}
The Beijing Rural Commercial Bank applied quantum encryption technology to its metro area ring network. This enabled secure transmission of office, production, and disaster recovery data among various locations, marking the first use of quantum communication technology in a metro area optical fiber ring network within the financial industry.

\item \textbf{World’s First Commercial Ultra-long Distance Co-propagation of QKD and Terabit Classical Optical Data Channels}
The China Telecom Beijing Institute, with QuantumCTek, Fiber Home, and ZTE Anhui Wan Tong, released the test results for the world’s first commercial ultra-long distance co-propagation of QKD and large-capacity classical optical data channels. This test utilized a commercial QKD system and an 8Tbps large capacity dense wavelength division multiplexing (DWDM) system, achieving QKD transmission over 100 kilometers without the use of trusted nodes.

\end{enumerate}

\section*{Declarations}\label{declarations}
    \subsection*{Availability of data and materials}
Data sharing does not apply to this article as no datasets were generated or analysed during the current study. The analysis of use case was performed entirely on the information and data available in the references listed.

\subsection*{Competing interests}
The authors declare that they have no competing interests.

\subsection*{Funding}
This work was in part funded by the Dutch Ministry of Economic Affairs and Climate Policy (EZK) as part of the Quantum Delta NL National Growthfunds on Quantum Technology and by the NWO NWA project FIQCS (NWA.1436.20.005).

\subsection*{Authors' contributions}
NA, BC and SD conducted the literature search for relevant use cases. NA, BC, SD, KH, FJW and SV contributed to the background and discussion of QKD and PQC. NA, BC, SD, KH, FJW and SV devised the analysis method. NA, SD, KH, FJW conducted the use case analysis. NA, BC, SD, KH and FJW wrote the manuscript. CO, SR, BS, ITM and SV reviewed the manuscript text. All authors contributed to final revision of the manuscript. All authors read and approved the final version of the manuscript.

\subsection*{Acknowledgements}
Not applicable.
 
\begin{CJK*}{UTF8}{min} \printbibliography
\end{CJK*}

\end{document}